\documentclass[a4paper,12pt]{article}

\usepackage[top=1.5in, bottom=1.5in, left=1.5in, right=1.5in]{geometry}

\usepackage{xcolor}
\usepackage{graphicx}
\usepackage{csvsimple}
\usepackage{booktabs}

\usepackage[round]{natbib} 
\usepackage{hyperref}
\usepackage{rotating}
\usepackage{threeparttable, caption}
\usepackage{subfig}

\graphicspath{ {Figures/} }

\usepackage{float}

\usepackage{hyperref}
\hypersetup{
    colorlinks=true, 
    linkcolor=blue, 
    filecolor=magenta,      
    urlcolor=blue,
    citecolor=blue,
}
 
\usepackage{amsmath,amsthm, amssymb}

\usepackage{setspace,lipsum}
\usepackage{setspace}
\usepackage{enumitem} 

\chardef\bslash=`\\ 

\theoremstyle{definition}

\begin{document}
\title{\vspace{-2.5cm} The Value Added of Machine Learning to Causal Inference: \\ Evidence from Revisited Studies\footnote{Baiardi acknowledges  support from EU Horizon 2020, Marie Skłodowska-Curie individual grant (No. 840319).  Naghi acknowledges support from EU Horizon 2020, Marie Skłodowska-Curie individual grant (No. 797286). Financial support from the United Nations Sustainable Development Funds is also gratefully acknowledged. We thank participants at the Machine Learning for Economics Workshop (at Barcelona GSE Summer Forum 2019) and at the Netherlands Econometrics Study Group Meeting 2020 for very helpful comments. Nadja van't Hoff and Christian Wirths provided excellent research assistance.}}

\author{Anna Baiardi\footnote{Department of Economics, Erasmus University and Tinbgergen Institute. Email: baiardi@ese.eur.nl.} \ and Andrea A. Naghi\footnote{Department of Econometrics,  Erasmus University and Tinbergen Institute.  Email: naghi@ese.eur.nl. }
}   

\date{December 2020}

\maketitle

\noindent \textbf{Abstract:} 
A new and rapidly growing econometric literature is making advances in the problem of using machine learning methods for causal inference questions. Yet, the empirical economics
literature has not started to fully exploit the strengths of these modern methods.
We revisit influential
empirical studies with causal machine learning methods and identify several advantages of using these techniques. We show that these advantages and their implications are empirically relevant and that the use of these methods can improve the credibility of causal analysis.

\noindent \textit{Keywords:} machine learning, causal inference, average treatment effects, heterogeneous treatment effects.
\\
\noindent \textit{J.E.L.} \textit{Classification}:  C01, C21, D04

\pagestyle{plain}
\pagenumbering{arabic}

\setcounter{page}{1}
\setcounter{secnumdepth}{5}
\setcounter{tocdepth}{5}


\newpage
\section{Introduction}

One of the key goals of empirical research in economics is to estimate the causal effect of a variable of interest on a targeted outcome.  To avoid biases in the coefficients of interest due to omitted variables, particularly in observational studies, it is often desirable to include a large number of controls. However, even when the number of raw covariates is relatively small, the inclusion of technical controls (e.g. dummy variables for geographical location, time periods, etc.), interactions and transformations can lead to settings in which the number of covariates is large relative to the sample size.

Machine learning (ML) methods can potentially be useful in such settings. However, standard ML prediction models are aimed for fundamentally different problems than most of the empirical work in economics. ML methods are designed and optimized for predicting the outcome in a test sample.\footnote{Note that by 'prediction' here, we do not mean 'forecasting'. Rather, we refer to a setting where we observe both the outcome and the features/covariates in a training sample and the aim is to predict the outcomes for each observation in an independent test sample, based on the actual values of the covariates in that test sample.} 
Thus, a model is selected by optimizing the goodness of fit on the held-out test set. In contrast, in empirical economic research, the goodness of fit of a model is oftentimes reduced when estimating a causal effect, and the predictive accuracy is sacrificed in order to learn more deeply about a fundamental relationship that can guide policy decisions and counterfactual predictions \citep{athey2019machine}. 
These fundamental differences will eventually generate biased estimates if \emph{standard} ML techniques, designed for prediction, are used in the context of causal inference.\footnote{The main underlying reason is that high dimensional regression adjustments such as lasso, ridge, elastic net etc., shrink the estimated effects by construction, and ignoring this shrinkage will lead to biased treatment effect estimates.} Nevertheless, a new and rapidly growing econometric literature is making advances in the problem of using ML methods for causal inference questions (see, e.g., \citealp{chernozhukov2018double, athey2018approximate, wager2018estimation, chernozhukov2018generic}). This literature brings in new insights and theoretical results that are novel for both the ML and the econometrics/statistics literature. Despite these advances, the empirical economics literature has not started yet to fully exploit the strengths of these new modern causal inference methods. 

The aim of this paper is to contribute to the general understanding of when and how ML methods add value to economic causal analysis. To this end, we revisit a number of influential papers with causal ML methods and compare the results with the traditional methods used in the original study. We highlight the relevance and additional gains that causal machine learning methods bring to the table relative to the standard econometric approaches. In our analysis, we focus on both the average treatment effect (ATE) and heterogeneous treatment effects (HTE). We provide evidence that causal ML methods can improve the credibility of causal analysis, and can help identify relevant heterogeneity in treatment effects that may be missed with traditional methods.

When interested in the ATE, we employ the double/debiased machine learning (DML) method of \citet{chernozhukov2017double}; when the focus is on heterogeneous treatment effects (HTE), we work with the random forest method of \citet{wager2018estimation}, and with  the generic machine learning method for heterogeneous treatment effects developed by \citet{chernozhukov2018generic}. These are newly developed causal machine learning methods for the estimation of the ATE and HTE with well established theoretical properties. We re-examine a set of relatively recent but influential studies that span a variety of topics in applied economics, published in the following journals: \textit{The Quarterly Journal of Economics, American Economic Journal: Macroeconomics, American Economic Journal: Applied Economics.} We choose papers for which the full replication data set is available either on the journal's website or on the authors' website. For the ATE, we revisit three observational studies: the study of  \citet{djankov2010effect} on the effect of corporate taxes on investment and entrepreneurship, the analysis of \citet{alesina2013origins} on the long-term effect of plough agriculture on gender norms,  and the paper by \citet{nunn2010structure} on the effect of skill-biased tariffs on long-term economic growth. For the HTE, we select one observational study and one randomized control trial: we extend the observational study by \citet{dellavigna2007fox}, which investigates the effect of Fox News on the Republican vote share, and the analysis by \citet{loyalka2019does} on the effect of a teacher training randomized intervention on student performance. All these papers include careful econometric analyses of the main research question and mechanisms, which we do not aim to re-examine in full. We instead focus on analyzing the main questions.

Our findings show important differences in the ATE and HTE estimates compared to the traditional methods, both in terms of size of the treatment effect estimates, and in terms of statistical significance. From our results, we derive four main observations about the reasons why causal machine learning methods are relevant for causal analysis and add value relative to the traditional methods. These observations are supported by the theoretical econometrics literature on causal ML (see, for example, \citealp{athey2018approximate, chernozhukov2018generic, wager2018estimation}).

Firstly, causal ML methods are powerful tools in using data to recover  complex  interactions among  variables  and  flexibly  estimate the relationship between the outcome,  the treatment indicator and covariates. This feature is key when drawing inference based on the assumption that the treatment is unconfounded as in the case of most of the revisited studies, since
this assumption is not testable. As some covariates can be correlated with both the
treatment variable and the outcome, failing to condition on all relevant confounders may
lead to biased estimates for the treatment effect. For example, for the effect of corporate taxes on investment and entrepreneurship, the original analysis in \citet{djankov2010effect} shows a negative and significant
effect of corporate taxes on investment and entrepreneurship, but the authors show that
these results do not survive when conditioning on all the potential controls at once. However, when
implementing DML, we obtain larger estimates compared to \citet{djankov2010effect},
which are often statistically significant. 
Similarly, our DML results for the effect of plough cultivation on gender roles suggest a larger effect of the plough compared to the findings in \citet{alesina2013origins}, when we use the instrumental variable strategy employed in the original analysis. Furthermore, our analysis of the effect of skill-biased tariffs on growth suggests a smaller effect compared to \citet{nunn2010structure}, which is often not statistically significant. We thus argue that the DML estimates are more robust to potential nonlinear confounders.\footnote{It is important to note here that the idea of estimating treatment effects without making parametric
assumptions about the way in which the covariates enter the equation has already been  considered in the semiparametric econometrics literature (see the review paper of \citealp{imbens2009recent}, and \citealp{imbens2015causal}) . However, in practice, these semiparametric
kernel methods quickly break down if they have to deal with more than a few
covariates.}

Secondly, causal ML methods allow for the inclusion of a large number of covariates, even when the sample size is relatively small, by assuming that the model is sparse, (i.e., only a small number of covariates
are relevant), and using regularized regressions. For instance, in the study by \citet{djankov2010effect} and in some of the specifications in \citet{alesina2013origins} and \citet{nunn2010structure}, the number of ``raw" covariates is large compared to the sample size, thus taking into account all possible nonlinear terms, such as interactions and transformations, would not be possible when using traditional methods. Indeed, only a limited set of prespecified nonlinear terms is included in \citet{alesina2013origins}, no nonlinear terms other than logarithms are considered in \citet{nunn2010structure}, and no nonlinear terms are included in \citet{djankov2010effect}. In contrast, by using the DML method we ensure that our results take into account all potentially relevant confounders at once, both linearly and nonlinearly.

Thirdly, the use of causal ML methods allows to implement \emph{systematic model selection}. ML methods search for the best functional forms by estimating and comparing a
wide range of alternative model specifications; the model selection is thus data-driven
and fully documented. 
For example, our results for the effect of corporate taxes, originally explored by \citet{djankov2010effect}, show that the data-driven model selection implemented by DML, which keeps a smaller set of influential confounding factors from among a large set of potential controls, leads to larger coefficients in absolute value and lower standard errors compared to OLS regressions where all the covariates are included. With the traditional approach to model selection, uncertainty about the correct specification of the model can lead to choices that are relatively \textit{ad hoc}; different specifications may
lead to different point estimates, which in turn may lead to different policy decisions. Moreover, we further illustrate how these methods are also very useful tools for \emph{supplementary analyses} or \emph{robustness checks}. Typically, supplementary analysis is performed by presenting a number of selected regression specifications, while the approach of causal ML methods is more systematic, and ensures that important transformations of covariates that are not considered relevant a priori are not missed. For instance, when revisiting the OLS robustness analysis of \citet{alesina2013origins}, the DML results show small and insignificant treatment effects, in contrast to the original robustness regressions. Furthermore, our analysis of \citet{nunn2010structure} shows that the main results are not robust to flexibly controlling for a data-driven function of the covariates.

Finally, causal machine learning methods prove to be very
useful when one is interested in estimating heterogeneous treatment effects. As
causal ML methods can handle many covariates potentially responsible for effect heterogeneity
in a systematic way, it is less likely that relevant heterogeneous effects will be missed, compared to manually modelling different interaction terms. This feature is exemplified by our analysis of the heterogeneous effects of Fox News on the Republican vote share first explored by \citet{dellavigna2007fox} and of the teacher training intervention studied by \citet{loyalka2019does}: our results reveal drivers of heterogeneity that were unexplored in the original analysis. 
In addition, note that causal ML methods tailored for estimating heterogeneous treatment effects provide valid confidence intervals in high dimensional settings, as opposed to traditional methods where standard $p$-values for single hypothesis testing are not reliable. This is due to the multiple hypothesis testing problem, which can occur when researchers search iteratively for treatment effect heterogeneity, over a large number of covariates.\footnote{While solutions have been proposed to correct for the issue of multiple hypothesis testing (for example, \citealp{list2016multiple}), when the number of covariates is large, the power of these approaches to detect heterogeneity is low \citep{athey2017state}.}$^{,}$\footnote{A related issue is the ex-post selection of significant heterogeneous effects. To avoid this problem, in randomized control trials researchers are often required to specify before the experiment which heterogeneous effects they are interested to look into, in order to avoid searching for, and only reporting, significant effects. However, this limits the ability of the researcher to find unexpected relevant heterogeneity. Causal ML methods ensure that relevant heterogeneity is not missed while also providing valid confidence intervals. In addition, in observational studies, where pre-analysis plans are not common practice, causal ML methods can be particularly useful.}

The \emph{econometric theory} literature on adapting standard machine learning techniques to causal inference  questions is by now fast growing. See for example \citet{chernozhukov2017double}, \citet{chernozhukov2018double}, \citet{athey2018approximate}, \citet{farrell2018deep} \citet{colangelo2020double} for the ATE; and \citet{athey2016recursive}, \citet{wager2018estimation}, \citet{athey2019estimating}, \cite{chernozhukov2018generic}, \citet{semenova2018orthogona} for the HTE. In the statistics literature, estimation of HTE with machine learning methods has been the focus in \citet{hill2011bayesian}, \citet{imai2013estimating}, \citet{su2009subgroup}, \citet{zeileis2008model}, among others. A few papers started employing the above mentioned methods in interesting \emph{early applications}. See for example, \citet{davis2017rethinking}, \citet{davis2017using}, \citet{knaus2020heterogeneous}, \citet{strittmatter2019value} and \citet{bertrand2017contemporaneous} for the causal random forest, and \citet{deryugina2019mortality} for the generic machine learning.

In what follows, we present our methodology and main findings on the ATE using double machine learning in Section \ref{sect:resultsATE}. The methodology and analysis of HTE via the causal random forest is summarized in Section \ref{sect:resultsHTE1}. Section \ref{sect:resultsHTE2} focuses on the methodology and analysis of HTE using the generic machine learning method. Finally Section \ref{sect:conclusion} concludes.

\section{Average Treatment Effects with Double Machine Learning}
\label{sect:resultsATE}

This section contains the analysis on the ATE for the effect of corporate taxes on investment and entrepreneurship  \citep{djankov2010effect}, the effect of plough agriculture on gender roles \citep{alesina2013origins}, and the effect of skill-biased tariffs on growth \citep{nunn2010structure}, using the double machine learning method \citep{chernozhukov2017double}.

\subsection{Methodology: Double Machine Learning}
\label{sect:DML}

The method is suitable in settings with a large number of covariates relative to the sample size (either because the number of raw covariates is large to begin with, or there is a large number of technical controls), where typical non-parametric kernel or spline methods break down.

The main model specification of the method, in the notation of \citet{chernozhukov2018double}, is the partially linear regression: 
\begin{equation}
  Y=D\theta_{0}+g_{0}(X) + U  
\end{equation}
\begin{equation}
     D=m_{0}(X)+V
\end{equation}
where $Y$ is the outcome, $D$ is the treatment variable of interest, $X$ is a (high-dimensional) vector of controls, and $U$ and $V$ are disturbances. Equation (1) is the main equation of interest and the parameter $\theta_{0}$ is the treatment effect we would like to estimate. In this model, $\theta_{0}$ quantifies the \emph{average treatment effect}. The second equation is not of direct interest, but it keeps track of the dependence of the treatment on confounders. The covariates are related to the treatment through the function $m_{0}(X)$ and to the outcome variable through the function $g_{0}(X)$. While $m_{0}(X)$ and $g_{0}(X)$ can be nonlinear, the treatment variable, $D$, enters the model linearly (and additively). In observational studies, the function $m_{0}$ is typically nonzero, which means that the treatment assignment is not random, but depends on the covariates. The partially linear regression model is also extended to a partially linear IV model to allow for endogenous treatment. We refer to this model as DML-IV. 

A first idea one might have for estimating $\theta_{0}$ with ML methods would be to use a predictive-based ML approach and predict $Y$ using $D$ and $X$ to obtain $D\hat{\theta}_{0}+\hat{g}_{0}(X)$. This can be done for example by an iterative method that alternates between estimating ${g}_{0}$ with some ML method and $\theta_{0}$ with OLS. While this 'naive' ML approach will have very good prediction performances, the iterative ML estimator will be heavily biased with a slower than $1/\sqrt{n}$ convergence rate. The primarily reason for this poor performance is the bias introduced by \emph{regularization}. In order to optimize prediction and avoid overfitting the data with complex functional forms, ML methods use regularization and shrink the less important coefficients towards zero. This reduces overfitting by decreasing the variance of the estimator but at the same time introduces bias. The bias in estimating $g_{0}$ transfers to the parameter of interest $\theta_{0}$. The issue is similar to the omitted variable bias.

To overcome regularization bias, \citet{chernozhukov2017double} propose `double machine learning'
i.e., solving two predictions problems instead of one. First, a ML model is fitted for $m_{0}$ in the treatment equation, and the effect of $X$ is partialed out from $D$ to get the residuals $\hat{V}=D-\hat{m}_{0}(X)$. Second, a ML method is fitted for $g_{0}$ in the outcome equation and the residuals $\hat{W}=Y-\hat{g}_{0}(X)$ are obtained.\footnote{The nuisance functions $m_{0}$ and $g_{0}$ can be estimated with a variety of ML methods such as: lasso, regression trees, random forest, boosting, neural networks, or hybrid methods.} Finally, the residuals $\hat{W}$ are regressed on the residuals $\hat{V}$ to obtained the `debiased' machine learning estimator, $\check{\theta}_{0}$. It can be shown that by orthogonalizing $D$ with respect to $X$ and eliminating the effect of confounders by subtracting an estimate of $g_{0}$, $\check{\theta}_{0}$ removes the effect of regularization bias.\footnote{This is because the scaled estimation error, $\sqrt{n}$($\check{\theta}_{0}-\theta_{0})$, contains now a term based on the product of two estimation errors (the estimation errors in $\hat{m}_{0}$ and in $\hat{g}_{0}$), which vanishes faster than the equivalent term obtained from using the naive estimator that depends on the estimation error of $\hat{g}_{0}$.}

However, $\check{\theta}_{0}$ is still subject to bias due to \emph{overfitting}. For instance, when $\hat{g}_{0}$ is overfit, it will mistake noise for signal and thus it will pick up some of the noise $U$ from the outcome equation. If $U$ and $V$ are correlated, the estimation error in $\hat{g}_{0}$ will be correlated with $V$. To break this correlation and avoid bias due to overfitting, one can rely on sample splitting. To this end, the data is partitioned into two subsamples: a main sample and an auxiliary sample. The ML models for the two nuisance functions $m_{0}$ and $g_{0}$ are fit on the auxiliary sample, while the residual on residual regression to obtain $\check{\theta}_{0}$ is fit on the main sample. 

A drawback of sample splitting is that the estimator of the parameter of interest $\theta_{0}$ is obtained using only the main sample, which can lead to loss of efficiency. However, one can switch the role of the main and auxiliary samples (procedure called \emph{cross-fitting}) and average the results, which will lead to a more efficient estimator. In addition, one can perform a  $K$-fold version of the cross-fitting procedure, where the size of each fold is $n/K$. Each sample partition or fold is successively taken as the main sample while the complement for each fold will be the auxiliary sample. One can take then the average of the estimates over the $K$-folds. To make the results robust to data partitioning, the splitting in folds procedure is performed $S$ times, and the final DML estimator is the mean (or median) over the splits. The median version is more robust to outliers and this is the one we use in the applications.

\subsection{Applications with Double Machine Learning}

\subsubsection{The Effect of Corporate Taxes on
Investment and Entrepreneurship}

\textbf{Description of Original Analysis.} The first paper that we revisit using causal machine learning methods investigates the relationship between corporate taxes on investment and entrepreneurship \citep{djankov2010effect}. This is an observational study that shows a negative effect of corporate taxes on investment and entrepreneurship, by estimating OLS country-level regressions with different measures of corporate tax rates for the year 2004. The sample includes a set of 50-85 countries, depending on the specification. In the original paper, four outcome variables are examined:  investment as a percentage of GDP, FDI as a percentage of GDP, business density per 100 people, and the average  entry  rate.   Three  measures  of  corporate  taxes are considered:  statutory corporate tax rates, actual first-year corporate income tax liability of a new company, and the tax rate which takes into account actual depreciation schedules going five years forward.

The original paper reports the results for several regression specifications with different sets of control variables, to account for potential confounders that correlate with corporate tax rates, and are also determinants of the outcomes.\footnote{The first set of controls includes measures of other taxes; the second set includes measures for the number of other tax payments made and for tax evasion; the third set includes measures for institutions; the fourth set includes measures of inflation. Section \ref{sect:taxes_paper} of the Appendix includes more details on the regressions estimated in \citet{djankov2010effect} and describes the control variables.} \citet{djankov2010effect} present regression results where the first  three  sets  of  covariates are added separately.
A final robustness check includes all control variables (12 in total) in the same regression. In the specifications which include only one set of controls at a time, the paper shows a negative and statistically significant effect of corporate taxes on entrepreneurship and investment. However, when adding all the controls, the relationship is still negative, but the coefficients are smaller in size and no longer statistically significant.

\textbf{DML Analysis.} We revisit the final robustness check of the paper, which includes all four sets of covariates at the same time, using the DML partially linear model.  
Table \ref{tab:tax} presents the results. Columns (1) to (7) display the DML point estimates for the effect of corporate taxes on investment and entrepreneurship, using different ML methods to estimate the nuisance functions. 
Further details on how the DML estimates are obtained, the methods used and the tuning parameters are described in Section \ref{sect:taxes_paper} of the Appendix.

\begin{table}[!t]
\caption{The Effect of Corporate Taxes on
Investment and Entrepreneurship}
\label{tab:tax}
\small
\centering
  \resizebox{\textwidth}{!}{\begin{tabular}{l c c c c c c c c}
    \toprule

  &  (1) & (2) & (3) & (4) & (5) & (6) & (7) & (8) \tabularnewline
     & Lasso & Reg. Tree & Boosting & Forest & Neural Net.  & Ensemble  & Best & OLS \tabularnewline   

  \cmidrule(lr){2-9}  

 & \multicolumn{8}{c}{\textit{Panel A: Investment 2003-2005}} \tabularnewline \tabularnewline
Statutory corporate tax rate & -0.074&	-0.069&	-0.068&	-0.07&	-0.056&	-0.066&	-0.071  &	-0.064 \tabularnewline

&(0.09) &	(0.072)&	(0.076)	&(0.087)&	(0.102)	&(0.087)&	(0.088)  &  (0.098)

 \tabularnewline
First-year effective tax rate &-0.114&	-0.129&	-0.154&	-0.144&	-0.122&	-0.13&	-0.133  &	-0.117 \tabularnewline
&(0.094)&	(0.087)	&(0.093)&	(0.096)&	(0.097)	&(0.092)&	(0.095) & (0.106)

 \tabularnewline
Five-year effective tax rate & -0.187&	-0.182&	-0.211&	-0.21&	-0.217&	-0.216&	-0.207& 	-0.189 \tabularnewline
 & (0.089)&	(0.089)&	(0.092)&	(0.097)&	(0.103)&	(0.095)&	(0.101) 
& (0.118) \tabularnewline

Observations & 61& 61&61 &61 &61 &61&61&61 \tabularnewline

\hline  
 & \multicolumn{8}{c}{\textit{Panel B: FDI 2003-2005}} 
 \tabularnewline \tabularnewline
Statutory corporate tax rate & -0.148&	-0.157	&-0.153&	-0.14&	-0.085&	-0.133&	-0.114  &	-0.030\tabularnewline

&(0.083)&	(0.086)	&(0.092)&	(0.094)&	(0.093)	&(0.088)&	(0.09) 
& (0.066) \tabularnewline
First-year effective tax rate & -0.141& 	-0.194&	-0.178&	-0.157&	-0.136&	-0.161&	-0.137 &	-0.1 \tabularnewline
&(0.091)& 	(0.081)& 	(0.081)& 	(0.074)& 	(0.078)& 	(0.08)& 	(0.079) 
& (0.071)\tabularnewline
Five-year effective tax rate & -0.147&	-0.177&	-0.167&	-0.165&	-0.139	&-0.157&	-0.14 &	-0.095 \tabularnewline
&(0.084)&	(0.073)	&(0.074)&	(0.077)&	(0.082)&	(0.077)&	(0.076) 
& (0.081) \tabularnewline 
Observations & 61& 61&61 &61 &61 &61&61&61 \tabularnewline

\hline  
 & \multicolumn{8}{c}{\textit{Panel C: Business density}} 
 \tabularnewline \tabularnewline
Statutory corporate tax rate &-0.062&	-0.092&	-0.069&	-0.07&	-0.056&	-0.066&	-0.06  &	-0.034\tabularnewline

&(0.066)&	(0.072)&	(0.061)	&(0.063)&	(0.077)&	(0.069)	&(0.064) 
& (0.083) \tabularnewline
First-year effective tax rate & -0.104&-0.156&	-0.124&	-0.122&	-0.105&	-0.114&	-0.1  &	-0.068 \tabularnewline
&(0.076)&	(0.082)	&(0.07)	&(0.069)&	(0.085)	&(0.072)&	(0.07) 
& (0.092)\tabularnewline
Five-year effective tax rate &-0.091&	-0.139&	-0.122&	-0.107&	-0.115&	-0.114&	-0.104  &	-0.070 \tabularnewline
&(0.076)&	(0.08)&	(0.071)	&(0.067)&	(0.087)&	(0.074)	&(0.075) 
& (0.103) \tabularnewline 
Observations & 60& 60&60 &60 &60 &60&60&60 \tabularnewline

 \hline  
  & \multicolumn{8}{c}{\textit{Panel D: Average entry rate 2000-2004}} \tabularnewline \tabularnewline
Statutory corporate tax rate & -0.112&	-0.147	&-0.141&	-0.127&	-0.067&	-0.112&	-0.106& 	-0.029 \tabularnewline

&(0.073)&	(0.068)&	(0.064)&	(0.065)&	(0.084)&	(0.067)&	(0.069) 
& (0.086) \tabularnewline
First-year effective tax rate & -0.130&	-0.144	&-0.143&	-0.125&	-0.131&	-0.126&	-0.117& 	-0.083\tabularnewline
&(0.072)&	(0.064)	&(0.065)&	(0.066)&	(0.086)&	(0.07)&	(0.072) 
& (0.094) \tabularnewline
Five-year effective tax rate & -0.154 &	-0.153 &	-0.164 &	-0.164 &	-0.191 &	-0.168 &	-0.167  &	-0.133\tabularnewline
&(0.084)&	(0.069)	&(0.07)	&(0.07)	&(0.091)&	(0.08)&	(0.077) 
&  (0.103) \tabularnewline
Observations & 50& 50&50 &50 &50 &50&50&50 \tabularnewline
\hline
Raw covariates &12&12 &12 &12 &12 &12&12&12 \tabularnewline
\bottomrule \tabularnewline

\end{tabular}}
\begin{tablenotes}
\footnotesize
\item \textit{Notes:} Analysis of Table 5D of \citet{djankov2010effect} using DML. Column 8 reports the original paper estimates. 
Standard errors are reported in parentheses. Standard errors adjusted for variability across splits using the median method are reported for the DML estimates. The number of covariates does not include the treatment variable.

\end{tablenotes}
\end{table}

We notice that all the DML point estimates have negative signs and generally similar magnitudes across the ML methods. Compared to the original paper results with the full set of covariates, reported in column (8), the magnitude of the DML coefficients is higher in absolute value, and the standard errors are lower in most regressions. Additionally, the results are statistically significant, at least at the 10\% level, in half (42 out of 84) of the regressions. The main difference between our and \citet{djankov2010effect}'s approach is that the original paper results are based on the assumption of linearity and additivity of the conditional expectations, while the DML method allows for a more flexible specification. Thus, our findings are more robust to potential nonlinear confounders compared to the original paper estimates. A researcher might be interested in investigating what are these nonlinear terms that make the estimates different. However, this can be a challenging task when ML methods (such as neural networks, hybrid methods etc.) are used to estimate the nuisance functions. What can potentially be done is analyzing the lasso coefficients that are not shrunk to zero and looking for nonlinearities among these. As an example, we show in Figure \ref{fig:taxlasso}  the most relevant among the nonlinear terms selected by the lasso, for one of the DML regressions reported in Table \ref{tab:tax}. Here, we note that several nonlinear terms appear in both the treatment nuisance function $\hat{m}(\cdot)$  and in the outcome nuisance function $\hat{g}(\cdot)$.\footnote{Further details about the lasso coefficients analysis are reported in Section \ref{sect:taxes_paper} of the Appendix.} This is suggestive of the fact that there are nonlinearities
that are correlated with both the treatment variable and the outcome. These were missed by the analysis in the original paper, and their omission could lead to biased coefficients of the corporate taxes variables. In this case, controlling for all relevant confounders strengthens the main results of the original analysis: in many cases the DML treatment effect estimates are larger in absolute value, and statistically significant.  

This empirical example is also useful to illustrate a typical trade-off that the applied researcher might face. On the one hand, the researcher wants to control for as many potential confounders as possible, in order to improve the credibility of the unconfoundedness assumption. On the other hand, naively controlling for a large set of covariates, especially when the sample size is small, can lead to imprecise estimates and larger standard errors. Notice that in this example, the authors implement a "kitchen sink"
regression and control for all the covariates at once, resulting in larger standard errors than the ones that we obtain. The DML method helps with this trade-off by improving the
credibility of the unconfoundedness assumption (as it captures more 
flexibly the effect of
confounders), but, at the same time,  it implements a data-driven variable selection
technique to keep a smaller set of influential confounding factors from among a large set of
potential controls, thus resulting in lower standard errors.

\subsubsection{The Effect of Plough Agriculture on Gender Roles}
\label{sect:ate_plough}

\textbf{Description of Original Analysis.} The study by \citet{alesina2013origins} examines the relationship between historical plough agriculture and gender roles. The mechanism is the following:  since the plough requires physical strength to be operated, in areas where plough agriculture was widespread, men had an advantage in agriculture compared to women. This would result in societies in which men worked in farming, whereas women's work would be performed mainly within the home. The division of labour by gender would translate into norms and cultural beliefs about the role of women in the society, which would still persist nowadays, even after societies have moved out of agriculture as the main economic activity. 

In the paper, the authors present results using country-level and individual-level regressions. We revisit the main question addressed in the original paper, focusing on the country-level results, as the majority of the regressions reported in the original paper are based on this data.
For the country-level baseline regressions, estimated with OLS, three contemporary outcome variables are examined as measures of gender roles: female labour force participation, the share of firms with a woman among its principal owners, and the proportion of seats held by women in the national parliament.
The treatment variable measures the share of individuals in each country whose ancestors practiced plough agriculture. 
The baseline regressions control for income, income squared, and for measures of the historical characteristics of the ethnicities living in a country. Continent fixed effects are added in some specifications.\footnote{More details on the regressions and the control variables from the study of \citet{alesina2013origins} are described in Section \ref{sect:plough_paper} of the Appendix.}

As mentioned by \citet{alesina2013origins}, concerns about potential endogeneity in the baseline regressions arise. It is possible that plough agriculture may have been more common in countries that had less equal
gender-role attitudes. This would cause the OLS estimates to be
biased away from zero. Moreover, plough agriculture may have been more likely in areas where economic development was historically higher. If historical and contemporary economic development are correlated, and more economically advanced countries tend to have higher female labour force participation and more equal gender roles, OLS estimates may be biased towards zero.
To tackle these issues, the following two solutions are offered in the paper.
First, motivated by the thought that the potential bias may be partly due to observable characteristics, a number of additional controls are included in the regressions. 
These include both historical and contemporary controls.\footnote{The additional controls are listed in Section \ref{sect:plough_paper} of the Appendix.} 
Second, the authors use an instrumental variable approach,
which exploits the fact that plough adoption is correlated with the suitability of the land for cereal crops that would benefit, and crops that would not benefit, from the plough. To this end, two instruments for plough adoption are constructed, based on the analysis by \citet{pryor1985invention}. The first is the suitability for "plough-positive" (i.e. which benefit most from the plough) cereal crops, and the second is the suitability for "plough-negative" (i.e. which benefit least from the plough) cereal crops.\footnote{See \citet{alesina2013origins} for details on the data used and how the instruments are constructed. The paper also shows that the two instruments, indeed, predict plough adoption.}

\textbf{DML Analysis.} In our analysis, we re-examine both the country-level OLS and IV regressions, applying the DML method. For the OLS analysis, we begin by estimating a DML partially linear model that only includes the baseline set of controls as raw covariates. We then revisit the robustness analysis of this specification, by including as raw covariates the largest set of controls used in the robustness checks (this corresponds to Table 7, column 8 of the original paper), to which we also add the continent fixed effects.\footnote{When revisiting the robustness analysis with DML, we include continent fixed effects, even though the original paper did not include them in their most complete robustness checks. As causal ML methods can handle a large number of covariates, we include all the covariate which were considered in the original paper, to ensure that all potential confounders are taken into account.} This amounts to a total of 36 raw covariates. 
For the IV analysis, as noted in \cite{alesina2013origins}, the main concern with the instrumental variable strategy is the possibility that suitable areas for different crops could be correlated with geographic characteristics that have an effect on gender norms through other channels, besides plough adoption (i.e., the exclusion restriction might not hold). Therefore, for the IV analysis, in addition to the baseline controls and in line with the original paper, we consider the geo-climatic characteristics the authors use in their IV robustness checks (Table A14 of the Online Appendix of the original paper). 
To these variables, we again add the continent fixed effects. 
Further details on how the DML estimates are obtained and the tuning choices are described in Section \ref{sect:plough_paper} of the Appendix. 

Table \ref{tab:ploughcountry} reports the results of the DML partially linear model that replicates the baseline regression. In accordance with the original paper, the treatment effect estimates are negative and statistically significant. They are also close to the original estimates (reproduced here for convenience in column 8 of Table \ref{tab:ploughcountry}), and reassuringly, fairly stable across the ML methods. We find however very different results when carrying out the robustness analysis of this baseline specification with the DML method. Panel A of Table \ref{tab:ploughcountry_ks} reports the results. While the effect is still negative, albeit much smaller in absolute value, statistically significance is now lost. Interestingly, when \citet{alesina2013origins}
include all covariates at once (the estimate is reproduced in the last column of our Table \ref{tab:ploughcountry_ks}),
the treatment effect becomes smaller
in absolute value, compared to when 
groups of covariates are added separately (see their Table 7, columns 1 to 7), or compared to the baseline specification (reproduced in column 8 of our Table \ref{tab:ploughcountry}). With DML, the treatment effect of interest does not only become smaller, but also statistically insignificant.

\begin{sidewaystable}[ph!]
\caption{On the origins of gender roles: Country-level estimates with full set of controls}
\label{tab:ploughcountry_ks}
\small
\centering
  \resizebox{\textwidth}{!}{ \begin{tabular}{l c c c c c c c c}
    \toprule

       &  (1) & (2) & (3) & (4) & (5) & (6) & (7) & (8)\tabularnewline
           & Lasso & Reg. Tree & Boosting & Forest & Neural Net.  & Ensemble  & Best & OLS\tabularnewline   
  \cmidrule(lr){2-9} 

 \tabularnewline
 & \multicolumn{8}{c}{\textit{Panel A: DML, Partially linear Model. Outcome: Female labour force participation}} \tabularnewline \tabularnewline
Traditional plough use & -5.922	&-6.744&	-6.445&	-5.812&	-5.682&	-5.925&	-5.224 & -9.234
 \tabularnewline
&(5.636)&	(5.063)&	(4.834)&	(4.742)&	(6.023)&	(4.998)
&(4.91)
 & (4.301) 

\tabularnewline
 Observations & 142 & 142 & 142 & 142 & 142 &  142&  142&  142\tabularnewline
  Raw covariates &36&36&36&36&36&36&36 & 30 \tabularnewline

\hline   \tabularnewline
     & Lasso & Reg. Tree & Boosting & Forest & Neural Net.  & Ensemble  & Best & 2SLS\tabularnewline    \cmidrule(lr){2-9} 
     \tabularnewline
 & \multicolumn{8}{c}{\textit{Panel B: DML-IV. Outcome: Female labour force participation}} 
 \tabularnewline \tabularnewline
Traditional plough use & -38.345 &	-36.85&	-39.429&	-36.961&	-20.725&	-33.645&	-38.712 & -28.516
 \tabularnewline
&(19.936)&	(18.996)&	(16.23)&	(14.247)&	(29.797)&	(21.486)&	(23.011) 
 & (7.559)
 \tabularnewline 
 Observations & 160& 160& 160& 160& 160& 160& 160 & 160 \tabularnewline
Raw covariates & 17& 17& 17& 17& 17& 17& 17 & 11\tabularnewline

\bottomrule \tabularnewline
\end{tabular}}
 \begin{tablenotes}
  \footnotesize
  \item  \textit{Notes:} Analysis of the main robustness checks of \citet{alesina2013origins} using DML. Column 8 reports the results of the most complete robustness checks for the OLS and IV specifications in the original paper. Standard errors adjusted for variability across splits using the median method are reported for the DML estimates. Robust standard errors are reported in column 8. The number of covariates does not include the treatment variable.
  \end{tablenotes}
\end{sidewaystable}

Our findings up to this point would lead us to (mistakenly) conclude that the negative effect of plough adoption on attitudes towards gender roles may not be as large as suggested by the original analysis, and that the effect is not statistically significant. However, our estimates from the DML partially linear model may still be subject to endogeneity. 
While flexibly controlling for a large number of covariates can account for the confounding effect of observed characteristics, the remaining concern is that plough adoption may be correlated with unobserved characteristics that also affect the outcome. 
The instrumental variable approach suggested by \citet{alesina2013origins} can alleviate this potential issue. 
We consider the same instruments as in the paper (described above) and we turn to re-evaluating the results by estimating a DML - IV model. Panel B of Table \ref{tab:ploughcountry_ks} reports the results. As in the original analysis, the estimated coefficients have a negative sign, and they are now statistically significant at the 10\% level for most of the ML methods, with the exception of neural networks and ensemble. It is interesting to note that the magnitude of the coefficients is larger than in the DML partially linear model (both baseline and extended). This is consistent with the original paper, which also finds that the IV coefficients are larger than the OLS estimates. It is worth to further notice that compared to the IV results of the original paper, our DML - IV findings suggest an even larger effect of the plough adoption on female labour force participation. We attribute this to causal machine learning methods being able to control for a large number of covariates in a more flexible way.\footnote{As explained above, our DML specification differ from the original paper's robustness analysis because it considers nonlinearities and it includes continent fixed effects. Therefore, the differences between the DML and the original estimates could, in principle, be driven by the continent fixed effects, and not by the nonlinearities. The original paper shows that adding the continent fixed effects to the baseline specification leads to very small changes in the OLS estimates (see Table 4 in the original paper), while it results in larger changes in the IV case (see Table 8 in the original paper). However, even in the IV case, including the continent fixed effects only increases the absolute size of the plough coefficient by 3-4 percentage points, while the DML coefficients exceed the OLS and 2SLS estimates by more than double that amount (with the except of the neural network and ensemble estimates). Thus, we conclude that allowing for a more flexible nuisance function is likely to be driving at least part of the differences between the DML and the 2SLS (and OLS) estimates.}
Overall, when looking at both the robustness analysis and the IV analysis and comparing them to the baseline results, we notice that our estimates move in the same direction as the original paper estimates, but our estimates move even more, supporting the idea that DML controls more flexibly for relevant covariates.

This empirical example is a good illustration to show the gains from combining modern ML tools with quasi-experimental methods such as  instrumental variables. While causal ML methods can make the unconfoundedness assumption more plausible by flexibly controlling for observed confounders, they cannot account for unobserved confounders. In such settings, the researcher could combine causal ML methods with quasi-experimental methods such as IV, which potentially overcomes biases caused by unobserved factors. Integrating the two methods could provide powerful tools for the researcher's toolkit.

Furthermore, this empirical paper illustrates how causal machine learning methods can serve as useful tools for the empirical researcher to perform supplementary analyses. In order to support the credibility of the empirical evidence, researchers typically report a number of different model specifications  and  evaluate  the  sensitivity  of  estimates  to  these alternatives -- similar to the above-mentioned robustness checks performed in the original paper. The usual approach to evaluating the variability of estimates to different model specifications can be somewhat ad-hoc and not a systematic way of implementing sensitivity analysis. In addition, relevant covariates or interactions of these covariates which are not considered important a priori by the researcher might be missed. Instead, causal machine learning methods use systematic algorithms that compare a wide range of model specifications for the nuisance functions and choose the one that best fits the data. This makes them more robust methods for sensitivity analyses than the current practice in the literature. Indeed, the example discussed here shows that the robustness analysis performed with DML suggest different conclusions compared to the original paper's robustness checks. Thus, we view causal machine learning methods as promising tools for sensitivity analysis in empirical work.

\subsubsection{The Effect of Skill-Biased Tariffs on Growth}

\textbf{Description of Original Analysis.} The study by \citet{nunn2010structure} investigates the relationship between skill-biased tariffs, i.e., a tariff structure that disproportionately favours skill-intensive industries, and long-term economic growth. The authors develop a theoretical framework based on \citet{grossman1991innovation} that shows how tariffs that focus on skill-intensive industries can lead to a disproportional expansion of skill-intensive industries, which then leads to higher long-term growth. Furthermore, using both cross-country and industry level data, the paper provides evidence of a positive relationship between the two variables, and delves into the mechanisms of this relationship. The findings suggest that the mechanisms from the theoretical framework can explain only part of the total correlation between skill-biased tariffs and growth.
The paper attributes the remaining part of the correlation to the endogeneity of skill-biased tariffs, and in particular to the relationship between institutions and the skill-bias of tariffs: countries with good institutions tend to protect more skill-intensive industries. 

In \citet{nunn2010structure}, three measures of the skill-bias of tariffs in the initial time period are used:\footnote{The initial time period is 1972 for 21 countries, 1980–83 for 30 countries and 1985-87 for 12 countries. The end period is 2000 for most countries, except for 3 of them, for which data ends in 1996. See \citet{nunn2010structure}, Table 1 for a list of the countries included and the respective time periods.} the correlation between the industry tariffs and the industry's skill-intensity, and two measures based on the difference between the log average tariffs in skill-intensive industries and log average tariffs in unskilled-intensive industries, which use different cut-off values for industry skill-intensity. In the country-level estimates, the outcome is log annual per capita GDP growth, and the regressions include a set of control variables.\footnote{Further details on the regressions estimated by \citet{nunn2007relationship} and on the control variables are described in Section \ref{sect:tariff_paper} of the Appendix.} The country-level regressions includes 63 observations.

For the industry-level estimates, the outcome variable is the average annual log change in industry output in each country, and the regressions include all the controls that appear in the country-level regressions, plus industry fixed effects. These regressions include 1004 data data points for 59 countries. An additional variable (the initial industry tariff) is included in some specifications to capture a potential mechanism: skill-biased tariffs can shift resources towards skill-intensive industries that generate  positive  externalities,  thus  leading  to  higher  long-term  growth. Thus, industries that have higher initial tariffs should have higher long-run output. If this channel can explain the effect of skill-bias on growth, the coefficient of the skill-bias of tariffs would decrease in size when this variable is included in the regression.

\begin{sidewaystable}[ph!]
\caption{The Structure of Tariffs and Long-Term Growth: Country-level estimates}
\label{tab:tariff1}
\small
\centering
  \resizebox{\textwidth}{!}{ \begin{tabular}{l c c c c c c c c}
    \toprule

   & (1) & (2)  & (3) & (4) & (5) & (6)  & (7) & (8)\tabularnewline
     & Lasso     &  Reg. Tree     &  Boosting     &  Forest     &  Neural Net.     &  Ensemble     &  Best & OLS  \tabularnewline   
  \cmidrule(lr){2-9} 

 \tabularnewline
 & \multicolumn{8}{c}{\textit{Panel A: Skill tariff correlation}} \tabularnewline \tabularnewline
Skill tariff correlation &  0.018&0.016  &0.016  & 0.015 & 0.014 & 0.017 & 0.016 & 0.035 \tabularnewline
&(0.010) &(0.011) &(0.012) &(0.011) &(0.013) &(0.012)&(0.011) &(0.01) 
\tabularnewline

 \tabularnewline
\hline \tabularnewline
 & \multicolumn{8}{c}{\textit{Panel B: Tariff differential (low cut-off)}} 
 \tabularnewline \tabularnewline
Tariff differential (low cut-off) & 0.009 & 0.006 & 0.007 &0.008  &0.013  &0.009  &0.008  & 0.016 \tabularnewline
& (0.005) &(0.005) &(0.005) &(0.005) &(0.008) &(0.006)&(0.005) &(0.005) \tabularnewline 

 \tabularnewline
 \hline \tabularnewline
  & \multicolumn{8}{c}{\textit{Panel C:Tariff differential (high cut-off)}} \tabularnewline\tabularnewline
Tariff differential (high cut-off) & 0.011 & 0.008 & 0.009 &0.009  &0.005  &0.009  & 0.009 & 0.02\tabularnewline
  & (0.005) &(0.005) &(0.006) &(0.006) &(0.008) &(0.006)&(0.006) &(0.004) \tabularnewline \tabularnewline
\hline 
   Observations &63 & 63 &63  &63  & 63 & 63 &63  &63 \tabularnewline 

Raw covariates & 17 &  17 &  17 &  17 &  17 &   17&  17 &  17\tabularnewline
\bottomrule \tabularnewline

\end{tabular}}
 \begin{tablenotes}
  \footnotesize
  \item  \textit{Notes:} Analysis of Table 4 (columns 1, 2, 4) of \citet{nunn2010structure} using DML. Column (8) reports the original paper estimates. Standard errors are reported in parentheses. Standard errors adjusted for variability across splits using the median method are reported for the DML estimates. The number of covariates does not include the treatment variable.
  \end{tablenotes}
\end{sidewaystable}

\textbf{DML Analysis.} We revisit the country and industry-level regressions reported in Tables 4 (columns 1, 2 and 4), Table 5 (columns 1, 2 and 4) and Table 6 (columns 1, 3 and 7) of \citet{nunn2010structure}. 
Further details on how the DML estimates are obtained and on the tuning parameter values are reported in Section \ref{sect:tariff_paper} of the Appendix.

Table \ref{tab:tariff1} shows the results of the DML partially linear model using country-level data.  
The DML treatment effect estimates are considerably smaller than the original paper's across all ML methods and across the three different treatment variables. Moreover, the estimated effects are not statistically significant, except the coefficients estimated using the lasso, which are significant at the 10\% level. Additionally, we report the DML results using the industry-level data set (Table \ref{tab:tariff2} and Table \ref{tab:tariff3} show the results with and without including the initial industry tariff respectively). Similarly to the country-level estimates, the industry-level estimates are not statistically significant across all methods, except for the boosting estimates, which are significant at the 10\% level.

Overall, the DML results suggest that the correlation between skill-biased tariffs and long-term economic growth is not robust to controlling for an unknown function of the average tariff level, country characteristics, initial production structure, cohort and region fixed effects. Indeed, the fact that the DML estimates are insignificant points to the presence of nonlinear confounding effects that are not accurately captured by the OLS regressions. 

It is worth noting here that the original paper attributes most of the correlation found between the treatment variables and long-term growth to the endogeneity of the skill-biased tariff variables, arising from the fact that skill-biased tariffs are more likely in countries with better institutions. Interestingly, in this example the country-level DML estimates are in line with the notion that the direct effect of the skill bias of tariffs is smaller than what is estimated by the OLS regressions. Finally, our results only concern the relationship between skill-biased tariffs and long-run economic growth, and not the relationship between skill-biased tariffs and institutions, or between institutions and long-run growth, which are examined in the original paper. Thus, our findings are consistent with the alternative mechanism described in \citet{nunn2010structure}, i.e. the existence of a causal relationship between institutions and economic growth.

\section{Heterogeneous Treatment Effects with \\ Causal Random Forest}
\label{sect:resultsHTE1}

This section focuses on the analysis of HTE for the effect of Fox News on Republican voting \citep{dellavigna2007fox} using the causal random forest method \citep{wager2018estimation}.

\subsection{Methodology: Causal Random Forest}

The causal random forest method is an adaptation of the original random forest for prediction, introduced by \citet{breiman2001random}, to the problem of causal inference. In this section, we start by briefly presenting the general idea of standard regression trees used for prediction, after which we describe how causal trees and causal random forests work.

The idea of \emph{regression trees} is to partition (or split) the data into groups based on the values of the covariates. The groups that are eventually obtained  are referred to as leaves. First, one starts with the whole data set as one group. Then, for each value of each covariate, the regression tree algorithm forms candidate splits, by placing all observations that have a covariate value that is lower than than the current value in the left leaf, and all observation for which their covariate value is greater than the current value in the right leaf. Among all these candidate splits, the one that is implemented is the one that minimizes an in-sample criterion function, such as the mean squared error (MSE) of the outcome variable within a leaf.\footnote{This mean squared error is computed as the sum of the squared differences between the outcomes of each unit within a leaf and the mean of these units in the leaf.} For each of the two new leaves, the algorithm repeats the procedure until a stopping rule\footnote{The stopping criteria can be for example: a pre-specified maximum number of leaves, the iteration when the minimizing split gives a covariate over which the observations have been already split by, or the iteration when the  proposed split does not decrease the mean squared error any further.} is reached, resulting in a tree-format partition of the data. Using the terminal leaves,  when the purpose is prediction, the outcome variables of  out-of-sample observations can be predicted by determining which terminal leaf a new observation belongs to, based on the values of the covariates, and assigning as its predicted outcome the mean of the outcomes in that leaf. 

Next, we turn to the \emph{causal random forest} method of \citet{wager2018estimation} which  builds on the \emph{causal tree} method of \citet{athey2016recursive}. For the causal tree, first, a percentage $p$ from the sample $N$ is drawn without replacement. Then, the subsample $n=p*N$ is further randomly split in half to form a training sample $n_{tr}$ and an estimation sample $n_{e}$. Using only the training sample $n_{tr}$, for each value of each covariate candidate splits are formed and a regression tree as described above is constructed. The key difference in the causal case compared to the prediction case is the objective function that is optimized when determining the split to be implemented.

Due to the fundamental problem of causal inference, directly training machine learning methods on the difference $Y_{i}(1)-Y_{i}(0)$, i.e., the difference of the outcomes that observation $i$ would have experienced with and without the treatment, is not possible, as we do not observe both outcomes for any individual unit. Thus, instead of minimizing an infeasible MSE, \citet{athey2016recursive} propose to maximize a criterion function that rewards a split that increases the variance of treatment effects across leaves and penalizes a split that increases within-leaf variance. The goal is to accurately estimate treatment effects within leaves, while preserving heterogeneity across leaves. The split is performed if it increases the criterion function, compared to no split. When no more splits can be done, the tree constructed based on the first subsample is defined.

The subsequent step involves turning to the estimation sample $n_{e}$, and based on the covariates, sorting each observation in this sample into the same tree. Using only the estimation sample, the treatment effect in each leaf is computed as $\hat{\tau}_l=\bar{y}_{lt}-\bar{y}_{lc}$ i.e., the mean outcome difference between treated ($t$) and control ($c$) observations within a leaf ($l$). The final step consists in returning to the full sample of $N$ observations, examining to which leaf each observation belongs based on the values of their covariates, and assigning  that leaf's treatment effect as the predicted treatment effect of the observation. Given that estimates from a single tree can have a high variance, the whole algorithm described above is repeated for a number of $B$ subsamples on which a number of $B$ trees are obtained that eventually form a causal random forest. The predicted treatment effect for each unit will be the average of predictions for that particular observation across the trees.

Notice that independent samples are used for: i) growing the tree (splitting the data), and ii) estimating treatment effects within each leaf of the tree. This property is called \emph{honesty}. Honesty leads to two desirable characteristics: it reduces bias from overfitting, and it makes the inference valid, since the asymptotic properties of the treatment effect estimates are the same as if the structure of the tree had been exogenously given.\footnote{Sample splitting, in general, can be inefficient as part of the data is not used. However, this loss of precision does not happen in the case of causal random forests. This is because although no observation is allowed to be used within the same tree for both partitioning the covariate space and estimation, when the data is subsampled and the forest is obtained based on many trees, each individual unit will appear in both the training sample and the estimation sample of some tree.}
\citet{wager2018estimation} establish consistency and the first asymptotic normality results for random forests which are then extended for the causal setting. For valid confidence intervals, a consistent estimator of the asymptotic variance is proposed, based on an infinitesimal jackknife for random forests. Further details regarding the tuning parameters of the causal random forest are provided in Section \ref{sect:fox_paper} of the Appendix.

\subsection{Application: The Effect of Fox News on the Republican Vote Share}

\textbf{Description of Original Analysis.} In this section we revisit and further analyze the study by \citet{dellavigna2007fox}. 
This paper examines the impact of media bias on voting outcomes. Specifically, it analyzes the impact of the entry of a conservative cable television channel, Fox News, on the Republican Party's vote share in the United States. To identify the causal effect of Fox News on voting, the authors investigate whether towns where Fox News became available between 1996 and 2000 experienced an increase in the vote share for the Republican Party in Presidential elections during the same time period. 
The estimation is performed on a data set at the town level, comprising information on 9256 towns. 

We consider the main outcome variable, i.e. the change in the vote share for the Republican party between 1996 and 2000. The treatment variable is a dummy indicating whether Fox News had become available between 1996 and 2000. To capture potential confounders, a number of control variables are included in the regressions.
\footnote{Further details on the regressions and on the control variables in \citet{dellavigna2007fox} are described in Section \ref{sect:fox_paper} of the Appendix.}

\citet{dellavigna2007fox} find a positive effect of Fox News on the Republican vote share. Moreover, they explore heterogeneity along a selected set of town characteristics: the number of available cable channels, the share of urban population, and whether the town is in a swing or Republican district. They do this by adding to the regression interaction effects of these covariates with the treatment variable.\footnote{The findings are reported in Table 6 of the original paper.}

\textbf{Causal Forest Analysis.} 
We perform the HTE analysis using the causal random forest method. 
Exploring heterogeneous effects is important for this study, in order to understand whether there are town or district characteristics that act as effect modifiers. While the average effects are informative for the impact of Fox News on the whole sample, it is often the case that treatment effects are not homogeneous. It is possible that the effect of Fox News was concentrated in some areas only. Understanding better the characteristics of the areas which saw the strongest and weakest responses can shed light on the mechanisms. The aim of this exercise is two-fold. First, we take an agnostic view about the nature of heterogeneity, and we investigate whether there are town or district characteristics which are treatment effect modifiers. Second, we examine whether the HTE analysis from the original paper matches the results from the causal ML methods.

We focus on one of the two preferred specifications from the original paper: the one that includes district fixed effects. We present results for two versions of the causal random forest, which account for district-level effects in different ways. In the first set of results, we include in the analysis dummy variables indicating the congressional district where the town is located. In the second set of results, we implement a cluster-robust version of the random forest developed by \citet{athey2019estimating}, where we treat each district as a separate cluster.  
The advantage of the cluster-robust causal forest is that it does not assume that clusters have an additive effect on the outcome. Further details on the clustered-robust causal forest and tuning parameter values used for the analysis are discussed in Section \ref{sect:fox_paper} of the Appendix.

\begin{table}[t!]
\caption{Fox News - Causal Forest: Average treatment effects and test for heterogeneity}
\label{tab:foxnews_het_test}
\small
\centering
  \resizebox{.8\textwidth}{!}{\begin{tabular}{l c c}
    \toprule 

   & (1) & (2) \tabularnewline
& District dummies & Cluster-robust \tabularnewline
\cmidrule(lr){2-3}

Fox News effect (ATE) &0.0065 & 0.0065
\tabularnewline
& (0.0016) & (.0027)\tabularnewline
Fox News effect above median & 0.013 & 0.0072
\tabularnewline
&(0.0024) & (0.0028) \tabularnewline
Fox News effect below median & -0.0033 & 0.0044\tabularnewline
& (0.0021) & (0.0048) \tabularnewline

95\% CI for the difference & (0.01009, 0.02255) & (-0.00806, 0.01374) \tabularnewline

Observations & 9256 & 9256
\tabularnewline

 \bottomrule \tabularnewline
 \end{tabular}}
 \begin{tablenotes}
  \footnotesize
  \item  \textit{Notes:} This table reports the estimated average treatment effect and a test for overall heterogeneity using the causal forest. Standard errors are reported in parentheses. ***, ** and * * indicate significance at the 1\%, 5\% and 10\% levels respectively.
\end{tablenotes}
\end{table}

We begin by discussing the average treatment effect. The results are presented in Table \ref{tab:foxnews_het_test}.  
As in the original analysis, we find a positive and significant effect of Fox News on the Republican vote share, both when including district dummies, and when implementing the clustered-robust causal forest;
however, the standard error in the clustered forest is larger. 
Our results suggest that in towns where Fox News became available the Republican party obtained a higher vote share by 0.65 percentage points on average, compared to towns where Fox News was not available. 
The ATE estimates are similar to the original paper estimates, which range between 0.004 and 0.007 (reported in Table 4 of \citealp{dellavigna2007fox}, columns 4-7). 

Next, we want to assess whether the causal forest can recover heterogeneity of treatment effects. As pointed out in \citet{athey2019estimating}, 
we can group observations according to whether their estimated out-of-bag conditional average treatment effect (CATE) is above or below the median CATE, and we can estimate the average treatment effect separately for these two subgroups. These are reported in Table \ref{tab:foxnews_het_test} as \textit{Fox News effect above median} and \textit{Fox News effect below median}. The difference between the two subgroup estimates is large when including district dummies, suggesting that there is potential for heterogeneity, and it is statistically significant, as indicated by the fact that the 95\% confidence interval for the difference between the two estimates does not contain zero (see column 1 of Table \ref{tab:foxnews_het_test}). However, the same heuristic test for the clustered-robust forest does not detect significant heterogeneity in the treatment effect. This could indicate that heterogeneity in the model with district dummy variables is overstated, because the dummy variables cannot appropriately capture the district-specific effects. The cluster-robust causal forest offers a more flexible way to capture district-specific effects, and may be more suitable in this case.\footnote{\citet{athey2019estimating} find a similar result in their application, when comparing the causal forest without clustering with the cluster-robust version.}

Although the results of the test for overall heterogeneity are mixed, it is still possible for heterogeneity to be present along some of the covariates. Hence, we investigate whether any of the included covariates are possible sources of heterogeneity. To do this, for each variable, we split the sample in two parts, based on whether the value of the covariate of interest is below and above the median, and we estimate the average treatment effect for the two subsamples.
Table \ref{tab:foxnews_het} reports the HTE results along the variables that appear to be significant determinants of heterogeneity in both specifications, while \ref{tab:foxnews_het2} and \ref{tab:foxnews_het3} report the results for the remaining variables. 
In addition, to gain further insight into which variables are more important for heterogeneity, we compute a measure of variable importance \citep{athey2019estimating}.\footnote{See Section \ref{sect:fox_paper} of the Appendix for details on how this measure is constructed.}   
Tables \ref{tab:foxnews_importance_1} and \ref{tab:foxnews_importance_2} report the variable importance measure for the covariates included in the district dummy variable specification and for the clustered-robust forest, respectively. We note that for both specifications, the variable importance measure is decreasing smoothly and we do not observe any variable that clearly stands out in terms of importance.

\begin{table}[t!]
\caption{Fox News - Causal Forest: HTE analysis}
\label{tab:foxnews_het}
\small
\centering
  \resizebox{\textwidth}{!}{\begin{tabular}{l c c c}
    \toprule 
       &  (1) & (2) & (3) \tabularnewline

  & CATE below median & CATE above median  & $p$-value difference \tabularnewline

\cmidrule(lr){2-4}

  & \multicolumn{3}{c}{\textit{Panel A: District dummies}} \tabularnewline 

Employment rate, diff. btw. 2000 and 1990 &  0.00928 &0.00064 & 0.00656 \tabularnewline 
&   (0.00244) & (0.00203) &  \tabularnewline
Share high school degree 2000 & 0.00805  & -7e-05  & 0.00884 \tabularnewline
&   (0.00226) &(0.00213)  &  \tabularnewline

Decile 10 in no. cable channels available & 0.00877  & -0.0044 & 6e-05 \tabularnewline
 &  (0.00192) & (0.00264) &   \tabularnewline

\midrule 

  & \multicolumn{3}{c}{\textit{Panel B: Cluster-robust}} \tabularnewline 
 Employment rate, diff. btw. 2000 and 1990 &0.00938  & 2e-04  &  0.06885  \tabularnewline
 &(0.00254) &  (0.00436) &  \tabularnewline
Share high school degree 2000 & 0.00859& -0.00179 & 0.05296 \tabularnewline
 & (0.00303) & (0.00442)&  \tabularnewline
Decile 10 in no. cable channels available  & 0.00857  & -0.00495  & 0.02033\tabularnewline
 &  (0.00289) & (0.00506) & \tabularnewline

 \bottomrule 

 \end{tabular}}
        \begin{tablenotes}
  \footnotesize
  \item  \textit{Notes:} This table reports the effect of Fox News on the Republican vote share for towns with values below (column 1) and above (column 2) the median of each variable.  Column 3 presents the $p$-value for the null of no difference between the estimates in columns 1 and 2. Standard errors are reported in parentheses. 
\end{tablenotes}
\end{table}

Our results in Table \ref{tab:foxnews_het}
show that three variables appear to be significant determinants of heterogeneity (at least at the 10\% level)  in both specifications: the change in employment between 1990 and 2000, the share of the population with education level equal to high school degree, and the 10th decile in number of cable channels available. We observe that the effect of Fox News on Republican voting is stronger in towns that experienced a smaller increase in the employment rate between 1990 and 2000. This finding may relate to the phenomenon of economic voting, i.e. the fact that voters tend to reward incumbents during periods of economic prosperity (e.g. \citealp{fair1978effect, kramer1971short, lewis2000economic, pissarides1980british}). Areas that experienced lower economic growth (and a smaller increase in employment) may have been more easily persuaded to vote Republican in 2000, since prior to the Presidential election of 2000 a Democratic President (Bill Clinton) had been in power for two consecutive mandates. Moreover, we observe a larger effect of Fox News in towns where the share of population with education level equal to high school degree is below median.  
We also find a larger positive effect of Fox News in towns where the 10th decile in the number of cable channels is below median, while the effect is negative and insignificant in towns where this variable is above median.\footnote{The median value for the 10th decile in number of cable channels is zero; hence, towns with value of this variable above median correspond to towns that are in the top decile in terms of number of cable channels available.}

Next, we investigate whether the findings regarding heterogeneity from the original paper are confirmed with the causal forest.
\citet{dellavigna2007fox} found a larger effect of Fox News on the Republican vote share in towns with a smaller number of cable channels available when including district fixed effects. While we do not observe significant heterogeneity along this variable, our results for the 10th decile in the number of cable channels are in line with the findings of the original analysis, and hence suggest that the effect of Fox News diminishes in the presence of higher competition in cable channels. It is also interesting to note that the number of cable channels emerges as the variable with the highest importance score in both specifications, which further points to the importance of this variable for heterogeneity. 
When investigating heterogeneity along the political orientation of the district, we confirm the findings of \citet{dellavigna2007fox}: we observe no significantly different effect for swing districts, and we obtained mixed results for Republican districts, as we find a significantly smaller effect of Fox News in Republican districts (at the 10\% level) when including district dummies, but not with the cluster-robust forest.\footnote{\citet{dellavigna2007fox} found mixed results for Republican districts in different specifications.}  However, in contrast to the original analysis, we do not find a significant difference in the effect of Fox News in rural versus urban towns, despite this being the only heterogeneity result that is robust in all specifications in \citet{dellavigna2007fox}.

In conclusion, our analysis of the HTE of Fox News on Republican voting confirms some of the findings from \citet{dellavigna2007fox}, namely the presence of heterogeneity along the number of cable channels and no robust heterogeneous effects for districts with different political orientations, but as opposed to the original paper it does not show different effects for urban and rural areas. The analysis with the causal forest further uncovers additional heterogeneity that was previously unexplored, such as a larger effect in towns that experienced a smaller increase in the employment rate, and a larger effect in towns with a lower share of population with high school degree. Finally, including district dummy variables results in the causal forest detecting more heterogeneity in treatment effects compared to the cluster-robust version, both when implementing the overall heterogeneity test and when analysing the HTE in terms of individual covariates. However, the model with district dummy variables could overstate the heterogeneity compared to the cluster-robust forest if the district dummies do not appropriately capture the district-specific effects. This points to the need of a more careful treatment of the issue of clustered observations when employing causal random forests for empirical applications \citep{athey2019estimating}.

\section{Heterogeneous Treatment Effects \\ with Generic Machine Learning}
\label{sect:resultsHTE2}

This section focuses on the analysis of HTE for the effect of a teacher training intervention \citep{loyalka2019does} using the generic machine learning method \citep{chernozhukov2018generic}.

\subsection{Methodology: Generic Machine Learning}
\label{sect:generic}

A different causal ML approach for HTE 
is the generic machine learning method of \citet{chernozhukov2018generic}. To make inference possible, the method does not focus directly on the HTEs, but on \emph{features} of HTEs such as: the best linear predictor of the heterogeneous effects (BLP), the group average treatment effects (GATES) sorted by the groups induced by machine learning proxies, and the average characteristics of the units in the most and least affected groups, or classification analysis (CLAN). The generic machine learning method is thus useful for empirical work as: (1) it allows detection of heterogeneity in the treatment effect, (2) computes the treatment effect for different groups of observations (such as least affected or most affected groups), and (3) describes which covariates are correlated the most with the heterogeneity.

The approach is based on random splitting of the data into an auxiliary and a main sample. The two samples are approximately equal in size. Based on the auxiliary sample, a ML estimator, called proxy predictor, is constructed for the conditional average treatment effect (CATE). 
Any generic ML method can be used for this approximation (e.g., elastic net, random forest, neural network, etc.). The proxy predictors are possibly biased and consistency is not required. We simply take them as approximations and use them to estimate and make inference on features of the CATE. Based on the main sample and the proxy predictors, we can compute the estimates of interest: BLP, GATES and CLAN, and then make inference relying on many splits of the data in auxiliary and main samples. 

We give a brief description on how the method works in practice. Let $Y$ be the outcome of interest, $D$ the  binary treatment variable, and $Z$ a vector of covariates. Define $b_{0}(Z)=E[Y(0)|Z]$, the baseline conditional average and $s_{0}(Z)=E[Y(1)|Z]-E[Y(0)|Z]$, the conditional average treatment effect (CATE). Using the auxiliary sample we obtain ML estimators (or proxy predictors) for the baseline conditional average and the conditional average treatment effect. As mentioned above, these are possibly biased predictors and consistency is not required. Then, for each unit in the main sample, we compute the predicted baseline effects, $B(Z)$ and the predicted treatment effects, $S(Z)$. Note that the predicted treatment effects, $S(Z)$, are obtained as the difference between the  predictions for the treatment group model and the control group model.
Following the notation from \citet{chernozhukov2018generic}, the BLP parameters are obtained using the main sample, by estimating the following regression by weighted OLS, with weights $1/(p(Z)(1-p(Z))$: 
\begin{equation}
\label{5}
  Y=\alpha'X_{1}+\beta_{1}(D-p(Z))+\beta_{2}(D-p(Z))(S(Z)-\overline{S(Z)})+\epsilon,
\end{equation}
where $X_{1}=[1, B(Z)]$, $p(Z)=P[D=1|Z]$ is the propensity score, and $\overline{S(Z)}$ is the average of the predicted treatment effect estimates on the main sample. The control $B(Z)$ is included to improve efficiency.
Note that the component $(D-p(Z))$ is part of the regressor $(D-p(Z))(S(Z)-\overline{S(Z)})$. Thus, it orthogonalizes this regressor to all other covariates that are functions of $Z$. The coefficient $\beta_{1}$ gives the average treatment effect, while $\beta_{2}$ quantifies how well the proxy predictor approximates the treatment heterogeneity. If $\beta_{2}$ is different from zero, it means that there exists heterogeneity in the treatment effects.

Once we obtain the predicted treatment effects, we can divide the observations from the main sample in group: $G_{1}, G_{2}, \dots, G_{K}$, based on their treatment effects. In our empirical applications, we choose $K=5$, such that group $G_{1}$ contains the  observations with the lowest 20$\%$ treatment effects and $G_{5}$ contains observations with the highest 20$\%$ treatment effects. Then, using again the main sample, we obtain the sorted group average treatment effects by estimating the weighted regression: \begin{equation}
  Y=\alpha'X_{1}+\sum_{k=1}^{K}\gamma_{k}(D-p(Z))\cdot 1(G_{k})+\nu,
\end{equation}
where $1(G_{k})$ is an indicator function for whether an observation is in group $k$, and where the weights are the same as in $(\ref{5})$. The parameters $\gamma_{k}$ give the average effect in each group (GATES). Also, if the difference $\gamma_{k}-\gamma_{1}$ is significantly different from zero, we again have evidence for treatment effect heterogeneity between the most affected and least affected groups. 

Lastly, we can analyze the properties or characteristics of the most affected and least affected groups, via Classification Analysis (CLAN).
Let $g(Y, Z)$ be a vector of characteristics of an observation.  We can compute average characteristics of the most affected and least affected group i.e.,  $\delta_{1}=E[g(Y,Z)|G_{1}]$ and $\delta_{2}=E[g(Y,Z)|G_{K}]$, the parameters of interest being averages of variables directly observed. Similarly to GATES, we can compute and make inference on the difference $\delta_{k}-\delta_{1}$.

\subsection{Application: The Effect of Teacher Training on Student Performance}

\textbf{Description of Original Analysis.} We reanalyze a large-scale randomized experiment that investigates the effect of a teacher professional development (PD) program in China on student achievement and on other student and teacher outcomes. The experiment was first studied by \citet{loyalka2019does}. Three hundred mathematics teachers, each employed in different schools across one province, took part in the intervention. The teachers were randomly assigned to one of the different treatment arms: PD only; PD plus a continuous follow-up with additional material and tasks for the trainees; PD plus an evaluation of the extent to which the teachers remembered the content of the training sessions; or no PD (control group).
The PD intervention consisted of lectures and discussions. 
  
Randomization was implemented at the school level, and in each school one teacher was nominated to participate in the intervention.  The main results are obtained by estimating a cross-sectional regression, where the treatment variable is a dummy indicating the treatment arm that the school was assigned to. The data was collected at three points in time: at baseline, midline and endline. Outcomes are measured at midline, or endline, and the main outcome of interest is student math achievement.\footnote{As  \citet{loyalka2019does} show similar results when estimating the impact of the intervention at midline or endline, we focus on the outcome variables measured at endline.} The control variables include student characteristics, teacher characteristics and class size.\footnote{Section \ref{sect:teacher_paper} of the Appendix describes the regressions and the control variables.}

The original paper finds no significant effect of the PD intervention on students’ achievement after one academic year, neither for the PD intervention alone, nor for the PD combined with the follow up and/or the evaluation treatments. The authors also do not find any effect on other outcomes, such as teacher knowledge or student motivation. 
The lack of effectiveness of the program is attributed to several factors: the content was too theoretical, the PD was delivered passively, and teachers could face constraints in the implementation of the suggested practices in the schools.
Furthermore, the paper analyzes heterogeneous treatment effects, by interacting the treatment variable with a number of student and teacher characteristics: student’s household wealth, baseline achievement level, the amount of training the teacher has received prior to the intervention, student and teacher gender, whether the teacher has a college degree and whether the teacher majored in math. The findings suggest that the effect of the treatment on students’ achievement can differ by teacher characteristic; however, no heterogeneous effects are found in terms of characteristics of students. 

\textbf{Generic ML Analysis.} We extend the analysis of HTE conducted in the original paper, by implementing the generic machine learning method developed by \citet{chernozhukov2018generic}.
Exploring heterogeneous treatment effects is particularly relevant for this intervention, because a small and insignificant estimate for the ATE could hide significant heterogeneity.
Our aim is to dig deeper into the analysis of heterogeneous treatment effects. First, we investigate whether there is significant heterogeneity in treatment effects; second, we analyze whether causal machine learning methods, by implementing a systematic search for heterogeneity across a large number of covariates, can offer additional insights about the characteristics of those who benefited from the program and those who did not, compared to the traditional methods used in the original paper.

In our analysis, we focus on the main outcome of interest, i.e. student math achievement. Since the results in the original paper are consistently close to zero when comparing the three different treatment arms with the control group, we choose to only analyze one of the treatment arms, corresponding to the PD intervention plus the evaluation.  
The sample that we use includes 10,006 students in 201 schools. We follow \citet{loyalka2019does} and cluster standard errors at the school level. In addition to the full set of controls included in the original paper, we also add to our analysis other variables that could be treatment effect modifiers: the baseline values of a number of student-level variables, plus variables indicating teachers behaviour in the classroom, evaluated by students at baseline.\footnote{These additional variables are described Section \ref{sect:teacher_paper} of the Appendix. In \citet{loyalka2019does}, the baseline value of the outcome variable is included as a control. Hence, the baseline characteristics described above are not included in all regressions in the original analysis. However, we consider these characteristics as potential drivers of heterogeneity; therefore, we include the baseline values of all available variables in our heterogeneity analysis.}

\begin{table}[t!]
\caption{Teacher Training - Generic Method: Best Linear Predictor}
\label{tab:teacher_blp}
\small
\centering
  \resizebox{.65\textwidth}{!}{\begin{tabular}{l c c }
    \toprule 
 &  (1) & (2) \tabularnewline
&  ATE ($\beta_1$) & HET ($\beta_2$) \tabularnewline
\cmidrule(lr){2-3}

 \vspace{2mm}
Estimate &0.002 & 0.651  \tabularnewline  
 \vspace{2mm}
90\% Confidence Interval &(-0.068,0.072)& (0.312,0.990)  \tabularnewline  
 \vspace{2mm}
$p$-value &  1.000 & 0.0003 \tabularnewline
Observations & 10006 & 10006
\tabularnewline
 \bottomrule 
 \end{tabular}}
        \begin{tablenotes}
  \footnotesize
  \item  \textit{Notes:} The estimates are obtained using neural network to produce the proxy predictor \textit{S(Z)}. The values reported correspond to the medians over 100 splits.
\end{tablenotes}
\end{table}

The generic method can be used in conjunction with a range of ML tools and \citet{chernozhukov2018generic} provide two measures (Best BLP and Best GATES) to compare the performance of the different ML methods used for the estimation of the proxy predictors. We consider the following methods: elastic net, neural network, and random forest. Based on the results of the Best BLP and Best GATES analysis, reported in Table \ref{tab:teacher_comparison} of the Appendix, we choose to further work with the neural network.\footnote{Further details on the Best BLP and GATES measures and on the tuning parameters used in this analysis are discussed in Section \ref{sect:teacher_paper}.}

\begin{figure}[ht!]
 \caption{Teacher Training - Generic Method: GATES}
 \label{fig:teacher_gates}
 \centering

\centering
  \includegraphics[width=.6\columnwidth]{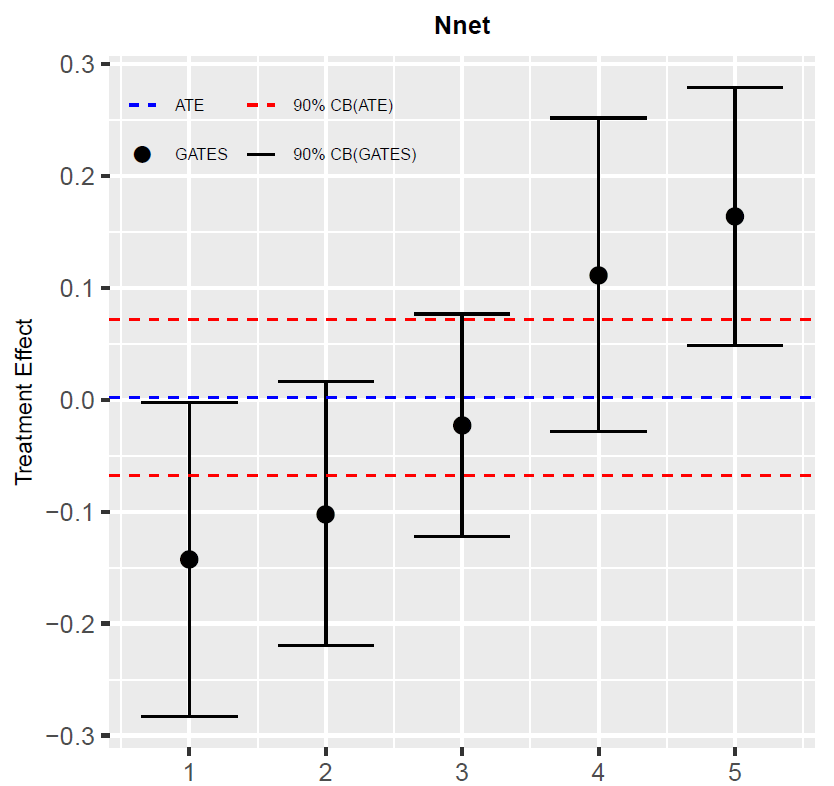}
   \begin{tablenotes}
  \footnotesize
    \item  \textit{Notes:}  The estimates are obtained using neural network to produce the proxy predictor \textit{S(Z)}. The point estimates and 90\% confidence intervals correspond to the medians over 100 splits.
    \end{tablenotes}
\end{figure}

We first analyze whether overall heterogeneity in treatment effects can be detected. We present results for the best linear predictor (BLP) of the CATE in Table \ref{tab:teacher_blp}. In line with the original paper, the estimated ATE, given by the coefficient $\beta_{1}$, is small (the estimated impact of the PD is 0.002 standard deviations) and not significantly different from zero. The estimated $\beta_{2}$ is instead large and significantly different from zero, which indicates that there is heterogeneity in treatment effects. Next, we estimate the group average treatment effects (GATES). We split the sample into five groups, based on the quintiles
of the ML proxy predictor \textit{S(Z)}. This analysis reveals further insights into the extent of heterogeneity. Table \ref{tab:teacher_gates} of the Appendix reports the GATE in the top and bottom quintile and shows that the GATE in the top quintile is positive, whereas for the bottom quintile the estimated GATE is negative. Both estimates are statistically significant at the 10\% level. The difference between the GATE for the top and the bottom quintile is significant, which confirms the presence of heterogeneity in treatment effects. Additionally, Figure \ref{fig:teacher_gates} reports the GATES estimate and the 90\% confidence interval for the five quintiles, as well as for the whole sample (the ATE is represented as a blue dashed line, and the confidence interval as two red dashed lines). Notice that for the three middle quintiles the effect of the teacher training intervention is not significantly different from zero.

We then turn to analysing the possible sources of heterogeneity, by implementing the Classification Analysis (CLAN). Thus, we analyze further the top and botton quintile in terms of ATE, for which the effect of the PD intervention is positive and negative respectively. In particular, we compare the student and teacher characteristics in the two groups. As a large number of covariates is available, we focus on the ten covariates for which the correlation with the proxy predictor, $S(Z)$, is highest, reported in Table \ref{tab:teacher_clan}. Table \ref{tab:teacher_clan2} in the Appendix shows the CLAN analysis for the remaining covariates.\footnote{Table \ref{tab:teacher_corr} reports the correlation for each of the covariates with $S(Z)$.}

\begin{table}[t!]
\caption{Teacher Training - Generic Method: Classification Analysis}
\label{tab:teacher_clan}
\small
\centering
  \resizebox{\textwidth}{!}{\begin{tabular}{l c c c}
    \toprule 
 &  (1) & (2) & (3) \tabularnewline
  & 20\% most affected & 20\% least affected & \textit{p}-value for the difference \tabularnewline
\cmidrule(lr){2-4}

Teacher college degree & 0.039 & 0.800  & 0.000\tabularnewline
& (0.019,0.059) &  (0.780,0.820)\tabularnewline
Teacher training hours &2.447 &1.684 & 0.000\tabularnewline
& (2.399,2.494) & (1.636,1.731)\tabularnewline
Teacher ranking &0.666 & 0.405  & 0.000\tabularnewline
& (0.635,0.697) & (0.374,0.437)\tabularnewline
Student age & 14.18 &13.73 & 0.000 \tabularnewline
& (14.11,14.25)  & (13.65,13.80)\tabularnewline
Teacher experience (years) &16.18& 13.16 & 0.000 \tabularnewline
& (15.60,16.76) & (12.58,13.74)\tabularnewline
Student female &0.417 &0.555  & 0.000\tabularnewline
& (0.385,0.449) & (0.523,0.587)\tabularnewline
Teacher age & 37.51& 35.01 & 0.000\tabularnewline
& (37.02,38.00) & (34.52,35.50)\tabularnewline
Student math score at baseline &-0.029 &0.169 & 0.005\tabularnewline
& (-0.088,0.031) & (0.110,0.229)\tabularnewline
Student baseline math anxiety &0.298 &-0.219&0.000   \tabularnewline
& (0.236,0.360) & (-0.281,-0.157)\tabularnewline
Class size & 52.87& 64.37 &0.000 \tabularnewline
& (51.82,53.93)&  (63.32,65.43)\tabularnewline
\bottomrule 

 \end{tabular}}
        \begin{tablenotes}
  \footnotesize
  \item  \textit{Notes:} This table shows the average value of the teacher and student characteristics for the most and least affected groups. The estimates are obtained using neural network to produce the proxy predictor \textit{S(Z)}.  90\% confidence intervals are reported in parenthesis. The variables \textit{Student math score at baseline} and \textit{Student baseline math anxiety} are normalized.  The values reported correspond to the medians over 100 splits. 

\end{tablenotes}
\end{table}

We start by analyzing the \emph{characteristics of the teachers} whose students belong to the least and most affected group.
Interestingly, the variable indicating whether the teacher has a college degree or not is the variable that is most correlated with the proxy predictor, and it was the only one among the variables tested which was found to be a treatment effect modifier across all treatment arms in the original paper.
The students in the top quintile are more likely to be taught by a teacher who does not have a college degree, compared to the students in the bottom quintile. This is consistent with the results from \citet{loyalka2019does}, who found that the intervention has a negative effect on students whose teachers have a college degree, but a positive effect on students whose teachers are less qualified. Hence, the PD may help teachers who are less qualified, but, for more qualified teachers, the benefits of the intervention on their students do not outweigh the negative effect of the teachers being absent from the classroom in order to participate in the intervention. Whether or not the teacher majored in math is found to be a potential driver of heterogeneity with the generic method (the results are reported in Table \ref{tab:teacher_clan2}), whereas in the original paper it was not found to be significant when considering the effect of the PD plus evaluation, which we focus on.\footnote{
When considering the PD plus follow-up, the authors find a significant negative effect on the scores of students whose teachers majored in math relative to the scores of those whose teachers did not.} 
The direction of the effect is consistent with what was found in the original analysis: the students in the top quintile are more likely to have been taught by a teacher who does not have a major in math, compared to the students in the bottom quintile. It is also interesting to note that the number of hours of training that the teacher received prior to the intervention, which is not found to be a determinant of heterogeneity in the original paper, is higher in the most affected group compared to the least affected group.\footnote{The variable indicating teacher training hours previous to the intervention is a categorical variable, based on the terciles of the continuous variable. As the continuous variable is not included in the replication data set of the original paper, for our analysis we use this categorical variable, which takes values 1 to 3, where 3 is the top tercile in the number of training hours.} This may reflect the fact that teachers who have had more training in the past may be able to better implement the suggestions from the PD intervention. Table \ref{tab:teacher_clan} shows that teacher rank, experience and age are higher in the most affected group compared to the least affected group. This is consistent with the existence of a similar mechanism: teachers who have more experience may be able to better implement the suggestions from the PD intervention. As the PD is mainly theoretical, having had other types of training, or having more experience, may be helpful for an effective implementation of the practices learned during the PD.

We then examine whether any of the \emph{student characteristics} are potential drivers of heterogeneity. In contrast to the findings in \citet{loyalka2019does}, who did not find heterogeneity in terms of student features, we find that students in the most affected group differ in terms of several characteristics compared to students in the least affected group. Among the most correlated with the heterogeneity score (listed in Table \ref{tab:teacher_clan}) are student age and gender: students in the most affected group are on average about half a year older than students in the least affected group, and the most affected group includes a larger share of male students. Additionally, students in the most affected group, on average, have a lower baseline math score, and tend to be more anxious about math. Thus, teacher PD could be more beneficial for weaker students, and for students who are more anxious about the subject. Finally, class size appears to be a possible determinant of heterogeneity: students who benefit more from the PD tend to be in smaller classes. This result suggests that in smaller classes it may be easier for teachers to implement some of the practices introduced during the PD training. For instance, \citet{loyalka2019does} mention having students work together in small groups as one of the techniques that were suggested in the PD; this technique is likely to be easier to implement in smaller classes. 

In conclusion, our analysis confirms the presence of heterogeneous effects of the teacher PD intervention, and uncovers a rich set of potential determinants of heterogeneity. 
With the GATES analysis, we are able to show that the achievement of students belonging to the bottom quintile is negatively affected by the intervention, while the achievement of students in the top quintile is positively affected by the intervention. This confirms what was suggested by \citet{loyalka2019does}: that there is a group of students who benefits from the intervention, and a group who does not. In addition, the GATES analysis shows that the effect is not significantly different from zero for the students belonging to the middle quintiles.
With the CLAN analysis, we can obtain a clearer picture of the characteristics of the groups who benefit and who do not from the intervention, compared to the original HTE analysis.
In line with \citet{loyalka2019does}, we find that teacher characteristics such as having a college degree or having a major in math are potential determinants of heterogeneity. However, our study uncovers additional differences (that were not identified in the original paper) between the least and the most affected groups, in terms of both teacher and student characteristics, such as teacher's  rank, experience, age and number of training hours, as well as student's gender, age, baseline math score, baseline math anxiety and class size.

\section{Conclusion}
\label{sect:conclusion}

Our main message is that appropriately combining predictive methods with causal questions adds value to traditional methods and should be more often explored in applied research. We argue that in each revisited study the researcher would have benefited from employing causal ML methods and would have gained additional insights not provided by standard causal inference tools. 

When the researcher works with an observational study and is interested in the ATE, causal machine learning methods can improve the credibility of causal analysis by making the unconfoundedness assumption more plausible -- as causal ML methods control for potential confounders in a more flexible way; implement a systematic model selection; and are robust approaches for sensitivity analysis.\footnote{Note then even if the empirical study is a randomized control trial and controlling for confounding factors is not necessarily needed, the use of causal machine learning methods can improve efficiency and provide more precise estimates with lower standard errors and tighter confidence intervals.} If the researcher is interested in HTE, causal machine learning methods can ensure that relevant heterogeneity and its determinants are not missed, or falsely discovered due to multiple hypothesis testing issues. Also, causal ML methods can be used to uncover heterogeneity ex-post, without being bound to explore HTE only for the specific subgroups indicated in the pre-analysis plan.

\bibliographystyle{plainnat}

\bibliography{CML_BN.bib} 

\begin{thebibliography}{38}
\providecommand{\natexlab}[1]{#1}
\providecommand{\url}[1]{\texttt{#1}}
\expandafter\ifx\csname urlstyle\endcsname\relax
  \providecommand{\doi}[1]{doi: #1}\else
  \providecommand{\doi}{doi: \begingroup \urlstyle{rm}\Url}\fi

\bibitem[Alesina et~al.(2013)Alesina, Giuliano, and Nunn]{alesina2013origins}
Alberto Alesina, Paola Giuliano, and Nathan Nunn.
\newblock On the origins of gender roles: Women and the plough.
\newblock \emph{The Quarterly Journal of Economics}, 128\penalty0 (2):\penalty0
  469--530, 2013.

\bibitem[Athey and Imbens(2019)]{athey2019machine}
S~Athey and GW~Imbens.
\newblock Machine learning methods economists should know about, arxiv.
\newblock 2019.

\bibitem[Athey and Imbens(2016)]{athey2016recursive}
Susan Athey and Guido Imbens.
\newblock Recursive partitioning for heterogeneous causal effects.
\newblock \emph{Proceedings of the National Academy of Sciences}, 113\penalty0
  (27):\penalty0 7353--7360, 2016.

\bibitem[Athey and Imbens(2017)]{athey2017state}
Susan Athey and Guido~W Imbens.
\newblock The state of applied econometrics: Causality and policy evaluation.
\newblock \emph{Journal of Economic Perspectives}, 31\penalty0 (2):\penalty0
  3--32, 2017.

\bibitem[Athey and Wager(2019)]{athey2019estimating}
Susan Athey and Stefan Wager.
\newblock Estimating treatment effects with causal forests: An application.
\newblock \emph{arXiv preprint arXiv:1902.07409}, 2019.

\bibitem[Athey et~al.(2018)Athey, Imbens, and Wager]{athey2018approximate}
Susan Athey, Guido~W Imbens, and Stefan Wager.
\newblock Approximate residual balancing: debiased inference of average
  treatment effects in high dimensions.
\newblock \emph{Journal of the Royal Statistical Society: Series B (Statistical
  Methodology)}, 80\penalty0 (4):\penalty0 597--623, 2018.

\bibitem[Bertrand et~al.(2017)Bertrand, Cr{\'e}pon, Marguerie, and
  Premand]{bertrand2017contemporaneous}
Marianne Bertrand, Bruno Cr{\'e}pon, Alicia Marguerie, and Patrick Premand.
\newblock Contemporaneous and post-program impacts of a public works program:
  Evidence from c{\^o}te d'ivoire.
\newblock {Working Paper}, 2017.

\bibitem[Breiman(2001)]{breiman2001random}
Leo Breiman.
\newblock Random forests.
\newblock \emph{Machine learning}, 45\penalty0 (1):\penalty0 5--32, 2001.

\bibitem[Chernozhukov et~al.(2017)Chernozhukov, Chetverikov, Demirer, Duflo,
  Hansen, and Newey]{chernozhukov2017double}
Victor Chernozhukov, Denis Chetverikov, Mert Demirer, Esther Duflo, Christian
  Hansen, and Whitney Newey.
\newblock Double/debiased/neyman machine learning of treatment effects.
\newblock \emph{American Economic Review}, 107\penalty0 (5):\penalty0 261--65,
  2017.

\bibitem[Chernozhukov et~al.(2018{\natexlab{a}})Chernozhukov, Chetverikov,
  Demirer, Duflo, Hansen, Newey, and Robins]{chernozhukov2018double}
Victor Chernozhukov, Denis Chetverikov, Mert Demirer, Esther Duflo, Christian
  Hansen, Whitney Newey, and James Robins.
\newblock Double/debiased machine learning for treatment and structural
  parameters.
\newblock \emph{The Econometrics Journal}, 21\penalty0 (1):\penalty0 C1--C68,
  2018{\natexlab{a}}.

\bibitem[Chernozhukov et~al.(2018{\natexlab{b}})Chernozhukov, Demirer, Duflo,
  and Fernandez-Val]{chernozhukov2018generic}
Victor Chernozhukov, Mert Demirer, Esther Duflo, and Ivan Fernandez-Val.
\newblock Generic machine learning inference on heterogenous treatment effects
  in randomized experiments.
\newblock {Working Paper}, National Bureau of Economic Research,
  2018{\natexlab{b}}.

\bibitem[Colangelo and Lee(2020)]{colangelo2020double}
Kyle Colangelo and Ying-Ying Lee.
\newblock Double debiased machine learning nonparametric inference with
  continuous treatments.
\newblock \emph{arXiv preprint arXiv:2004.03036}, 2020.

\bibitem[Davis and Heller(2017{\natexlab{a}})]{davis2017using}
Jonathan Davis and Sara~B Heller.
\newblock Using causal forests to predict treatment heterogeneity: An
  application to summer jobs.
\newblock \emph{American Economic Review}, 107\penalty0 (5):\penalty0 546--50,
  2017{\natexlab{a}}.

\bibitem[Davis and Heller(2017{\natexlab{b}})]{davis2017rethinking}
Jonathan~MV Davis and Sara~B Heller.
\newblock Rethinking the benefits of youth employment programs: The
  heterogeneous effects of summer jobs.
\newblock \emph{Review of Economics and Statistics}, pages 1--47,
  2017{\natexlab{b}}.

\bibitem[DellaVigna and Kaplan(2007)]{dellavigna2007fox}
Stefano DellaVigna and Ethan Kaplan.
\newblock The fox news effect: Media bias and voting.
\newblock \emph{The Quarterly Journal of Economics}, 122\penalty0 (3):\penalty0
  1187--1234, 2007.

\bibitem[Deryugina et~al.(2019)Deryugina, Heutel, Miller, Molitor, and
  Reif]{deryugina2019mortality}
Tatyana Deryugina, Garth Heutel, Nolan~H Miller, David Molitor, and Julian
  Reif.
\newblock The mortality and medical costs of air pollution: Evidence from
  changes in wind direction.
\newblock \emph{American Economic Review}, 109\penalty0 (12):\penalty0
  4178--4219, 2019.

\bibitem[Djankov et~al.(2010)Djankov, Ganser, McLiesh, Ramalho, and
  Shleifer]{djankov2010effect}
Simeon Djankov, Tim Ganser, Caralee McLiesh, Rita Ramalho, and Andrei Shleifer.
\newblock The effect of corporate taxes on investment and entrepreneurship.
\newblock \emph{American Economic Journal: Macroeconomics}, 2\penalty0
  (3):\penalty0 31--64, 2010.

\bibitem[Fair(1978)]{fair1978effect}
Ray~C Fair.
\newblock The effect of economic events on votes for president.
\newblock \emph{The Review of Economics and Statistics}, pages 159--173, 1978.

\bibitem[Farrell et~al.(2018)Farrell, Liang, and Misra]{farrell2018deep}
Max~H Farrell, Tengyuan Liang, and Sanjog Misra.
\newblock Deep neural networks for estimation and inference: Application to
  causal effects and other semiparametric estimands.
\newblock \emph{arXiv preprint arXiv:1809.09953}, 2018.

\bibitem[Grossman and Helpman(1991)]{grossman1991innovation}
Gene~M Grossman and Elhanan Helpman.
\newblock \emph{Innovation and growth in the global economy}.
\newblock MIT press, 1991.

\bibitem[Hill(2011)]{hill2011bayesian}
Jennifer~L Hill.
\newblock Bayesian nonparametric modeling for causal inference.
\newblock \emph{Journal of Computational and Graphical Statistics}, 20\penalty0
  (1):\penalty0 217--240, 2011.

\bibitem[Imai et~al.(2013)Imai, Ratkovic, et~al.]{imai2013estimating}
Kosuke Imai, Marc Ratkovic, et~al.
\newblock Estimating treatment effect heterogeneity in randomized program
  evaluation.
\newblock \emph{The Annals of Applied Statistics}, 7\penalty0 (1):\penalty0
  443--470, 2013.

\bibitem[Imbens and Rubin(2015)]{imbens2015causal}
Guido~W Imbens and Donald~B Rubin.
\newblock \emph{Causal inference in statistics, social, and biomedical
  sciences}.
\newblock Cambridge University Press, 2015.

\bibitem[Imbens and Wooldridge(2009)]{imbens2009recent}
Guido~W Imbens and Jeffrey~M Wooldridge.
\newblock Recent developments in the econometrics of program evaluation.
\newblock \emph{Journal of Economic Literature}, 47\penalty0 (1):\penalty0
  5--86, 2009.

\bibitem[Knaus et~al.(2020)Knaus, Lechner, and
  Strittmatter]{knaus2020heterogeneous}
Michael~C Knaus, Michael Lechner, and Anthony Strittmatter.
\newblock Heterogeneous employment effects of job search programmes: A machine
  learning approach.
\newblock \emph{Journal of Human Resources}, pages 0718--9615R1, 2020.

\bibitem[Kramer(1971)]{kramer1971short}
Gerald~H Kramer.
\newblock Short-term fluctuations in us voting behavior, 1896--1964.
\newblock \emph{American Political Science Review}, 65\penalty0 (1):\penalty0
  131--143, 1971.

\bibitem[Lewis-Beck and Stegmaier(2000)]{lewis2000economic}
Michael~S Lewis-Beck and Mary Stegmaier.
\newblock Economic determinants of electoral outcomes.
\newblock \emph{Annual Review of Political Science}, 3\penalty0 (1):\penalty0
  183--219, 2000.

\bibitem[List et~al.(2016)List, Shaikh, and Xu]{list2016multiple}
John~A List, Azeem~M Shaikh, and Yang Xu.
\newblock Multiple hypothesis testing in experimental economics.
\newblock \emph{Experimental Economics}, pages 1--21, 2016.

\bibitem[Loyalka et~al.(2019)Loyalka, Popova, Li, and Shi]{loyalka2019does}
Prashant Loyalka, Anna Popova, Guirong Li, and Zhaolei Shi.
\newblock Does teacher training actually work? evidence from a large-scale
  randomized evaluation of a national teacher training program.
\newblock \emph{American Economic Journal: Applied Economics}, 11\penalty0
  (3):\penalty0 128--54, 2019.

\bibitem[Nunn(2007)]{nunn2007relationship}
Nathan Nunn.
\newblock Relationship-specificity, incomplete contracts, and the pattern of
  trade.
\newblock \emph{The Quarterly Journal of Economics}, 122\penalty0 (2):\penalty0
  569--600, 2007.

\bibitem[Nunn and Trefler(2010)]{nunn2010structure}
Nathan Nunn and Daniel Trefler.
\newblock The structure of tariffs and long-term growth.
\newblock \emph{American Economic Journal: Macroeconomics}, 2\penalty0
  (4):\penalty0 158--94, 2010.

\bibitem[Pissarides(1980)]{pissarides1980british}
Christopher~A Pissarides.
\newblock British government popularity and economic performance.
\newblock \emph{The Economic Journal}, 90\penalty0 (359):\penalty0 569--581,
  1980.

\bibitem[Pryor(1985)]{pryor1985invention}
Frederic~L Pryor.
\newblock The invention of the plow.
\newblock \emph{Comparative Studies in Society and history}, 27\penalty0
  (4):\penalty0 727--743, 1985.

\bibitem[Semenova et~al.(2018)Semenova, Goldman, Chernozhukov, and
  Taddy]{semenova2018orthogona}
Vira Semenova, Matt Goldman, Victor Chernozhukov, and Matt Taddy.
\newblock Orthogonal machine learning for demand estimation: High dimensional
  causal inference in dynamic panels.
\newblock \emph{arXiv preprint arXiv:1712.09988}, 2018.

\bibitem[Strittmatter(2019)]{strittmatter2019value}
Anthony Strittmatter.
\newblock What is the value added by using causal machine learning methods in a
  welfare experiment evaluation?
\newblock {Working Paper}, 2019.

\bibitem[Su et~al.(2009)Su, Tsai, Wang, Nickerson, and Li]{su2009subgroup}
Xiaogang Su, Chih-Ling Tsai, Hansheng Wang, David~M Nickerson, and Bogong Li.
\newblock Subgroup analysis via recursive partitioning.
\newblock \emph{Journal of Machine Learning Research}, 10\penalty0
  (Feb):\penalty0 141--158, 2009.

\bibitem[Wager and Athey(2018)]{wager2018estimation}
Stefan Wager and Susan Athey.
\newblock Estimation and inference of heterogeneous treatment effects using
  random forests.
\newblock \emph{Journal of the American Statistical Association}, 113\penalty0
  (523):\penalty0 1228--1242, 2018.

\bibitem[Zeileis et~al.(2008)Zeileis, Hothorn, and Hornik]{zeileis2008model}
Achim Zeileis, Torsten Hothorn, and Kurt Hornik.
\newblock Model-based recursive partitioning.
\newblock \emph{Journal of Computational and Graphical Statistics}, 17\penalty0
  (2):\penalty0 492--514, 2008.

\end{thebibliography}

\clearpage

\appendix

\setcounter{table}{0}
\renewcommand\thetable{\thesection.\arabic{table}}

\section{Details on Revisited Studies and  Implementation of Causal ML Methods}\label{sect:revisited_papers}

\subsection{The Effect of Corporate Taxes on Investment and Entrepreneurship}\label{sect:taxes_paper}

\textbf{Details on the Original Analysis.} 
In \citet{djankov2010effect}, the baseline regression equation is the following:

    \begin{equation*}
    \centering
y_{c}=\alpha + \beta taxes_{c} +
\boldsymbol{X_{c}} \boldsymbol{\Gamma}  +  \epsilon_{c},
 \end{equation*}
where $c$ is an index for country. 
Four different outcome variables are examined:  investment as a percentage of GDP, FDI as a percentage of GDP, business density per 100 people, and the average  entry  rate  (measured  as  percentage).   Three  separate  measures  of  corporate  taxes are considered.  The first is the statutory corporate tax rates, which is the marginal tax rate on income a corporation has to pay assuming the highest tax bracket.  The second is the actual first-year corporate income tax liability of a new company, relative to pre-tax earnings. 
The third is the tax rate which takes into account actual depreciation schedules going five years forward. 

The term $\boldsymbol{X_{c}}$ denotes the control variables, aimed at capturing the effect of potential confounding factors.
This is an observational study, in which tax rates are not randomly assigned across countries. It is likely that there will be factors which are correlated with both the treatment (corporate tax rates), and with the outcomes (measures of entrepreneurship and investment). To deal with this issue, the effect of corporate taxes on the outcomes is estimated by adding several control variables to the regressions. 
The first set of control variables are measures of other taxes: the sum of other taxes payable in the first year of operation,  VAT tax, sales tax, and the highest national rate on personal income tax. The second set of covariates include the logarithm of the number of tax payments made (which is used as a measure of the burden of tax administration), an index of tax evasion, and the number of procedures to start a business. The third set of controls are institutional variables: a property rights index, an indicator of the rigidity of employment laws, a measure of a country's openness to trade, and the log of per capita GDP. The fourth set of covariates are measures of inflation: average inflation in the previous ten years, and seigniorage, which captures government reliance on printing money.

 \textbf{Details on the DML Analysis.} 
The results are based on 100 splits and 2 folds. The point estimates are calculated as the median across splits, and the standard errors are adjusted for the variability across sample splits using the median method, see \citet{chernozhukov2018double}.

We use two hybrid ML methods in our analysis. Ensemble is a weighted average of estimates from lasso, boosting, random forest and neural networks, the weights being chosen to give the
lowest average mean squared out-of-sample prediction error. Best
chooses the best method for estimating the nuisance functions in terms of the average
out-of-sample prediction performance among all the other methods. 

The lasso estimates are based on $\ell_{1}$-penalized regressions with the penalty parameter obtained through 10-fold cross-validation. As controls, for the lasso we consider the set of all raw covariates as well as  first-order interactions. For the rest of the ML methods, we consider the set of raw covariates as controls. The regression tree method fits a CART (classification and regression tree) tree with a penalty parameter (that restricts the tree from overfitting and makes sure that only splits that are considered “worthy” are implemented) obtained with 10-fold cross validation. The random forest estimates are obtained using 1000 trees, while the Boosting estimates are obtained with 1000 boosted regression trees. For the boosting, the minimum number of observations in trees' terminal nodes is set to 1 and the bag.fraction parameter is set to 0.5, except for Panel D of Table \ref{tab:tax}, where it is increased to 0.8. For the neural networks we used 2 neurons and a decay parameter of 0.01; the activation function is set to the linear function.\footnote{In general, the activation function can be set to the linear function for regression problems (when the outcome is continuous) and to the logistic function for classification problems (when the outcome is categorical).}

For the \textit{analysis of nonlinear terms with lasso}, we examine the estimated nuisance functions for the outcome \textit{average entry rate} and the treatment variable \textit{first-year effective tax rate}. 
In our analysis, for the estimation of the two nuisance functions, the lasso selects among the simple covariates, and their two-way interactions.\footnote{For the lasso estimation, depending on the application, other nonlinear terms could be added, such as the squares of the covariates, or three-way interactions.}
It is interesting to note that a large number of interaction terms is selected. Figure \ref{fig:taxlasso} depicts the seven largest interaction terms and their coefficients in  the treatment nuisance function $\hat{m}(\cdot)$  and in the outcome nuisance function $\hat{g}(\cdot)$.\footnote{
It is important to note here that we do not make inference using the lasso coefficients, but just analyze the magnitude of the coefficients, as a measure of the covariates' importance for predicting the outcome and the treatment variables.} 
The lasso coefficients are calculated as the median coefficients across splits.  Among these, some appear in both nuisance functions (the coefficients of the common terms are depicted in purple in Figure \ref{fig:taxlasso}).
A particular issue that appears with the lasso when the interest is on analyzing the interaction terms is worth mentioning here.
Since the lasso implements regularization by shrinking the smallest coefficients to zero, it is possible that interaction terms are included in the regression, but the coefficients of the raw covariates forming the interactions are shrunk to zero.
It is thus important to check whether the raw covariates forming these interactions also appear in the regression. If the coefficients on the raw covariates are shrunk, the coefficient of the 'pure' interaction terms might not be properly captured and the found interaction terms might actually reflect the effect of the raw covariates, diminishing the importance of our uncovered nonlinearities. Thus, when analyzing the relevance of the interaction terms, we are careful to only report the coefficients of the interactions for which both main effects are included in the lasso estimation. The lasso coefficients of all the raw covariates are reported in Table \ref{tab:taxes_lasso} of the Appendix.

\subsection{The Effect of Plough Agriculture on Gender Roles}
\label{sect:plough_paper}

\textbf{Details on the Original Analysis.} 
\citet{alesina2013origins} consider several empirical strategies and data sets. They start with OLS regressions performed using country-level and micro-level data. Then, to tackle possible endogeneity issues, the paper follows two approaches: first, several potential confounders are included in the regressions; second, an instrumental variable strategy is used.\footnote{The instrumental variable strategy is summarized in Section \ref{sect:ate_plough}.} Our focus is on the country-level regressions. 

The baseline OLS country-level results in the original analysis (reported in Table 4 of \citealp{alesina2013origins}) are obtained by estimating the following regression: 

    \begin{equation*}
          \centering
y_{c}=\alpha + \beta plough\,use_{c} +
 \boldsymbol{X_{c}^H}\boldsymbol{\Gamma} +  \boldsymbol{X_{c}^C}\boldsymbol{\Pi} +  \epsilon_{c},
\end{equation*}
where $c$ stands for country. In the paper, three outcome variables are examined as measures of gender roles: female labour force participation, attitudes about women's work, and attitudes about women as leaders. 
The first outcome variable is an indicator variable that equals one if the individual is in the labor force in 2000; the second is the share of firms with a woman among its principal owners in the period 2003-2010; finally, the third is the proportion of seats held by women in the national parliament in 2000.
The treatment variable, $plough\,use_c$, is calculated as the estimated proportion of individuals living in a country with ancestors that used the plough in pre-industrial agriculture. The vector $X_{c}^H$ includes historical ethnographic variables at the country level. These controls capture the historical characteristics of ethnicities living in a country, and they are meant to account for differences between ethnicities that historically adopted the plough and those that did not. They include: ancestral suitability for agriculture,
fraction of ancestral land that was tropical or subtropical, ancestral domestication of large animals, ancestral settlement patterns, and ancestral political complexity. The vector $X_{c}^C$ denotes contemporary country-level controls: natural log of real per capita GDP, and its square. These are included as the level of economic development is believed to have an impact on female labour force participation, and the square of per capita GDP is intended to capture the observed U-shaped relation between the two variables. Continent fixed effects are also added in some specifications.

The extended set of controls includes additional historical and contemporary controls.
Just as with the baseline controls, the additional historical controls are measures of the characteristics of the ancestors of the current population living in a country. These are: the intensity of agriculture; the proportion of subsistence provided by hunting and by the herding of large animals; the fraction of countries'
ancestors without land inheritance rules, with patrilocal
post-marital residence rules, and with matrilocal post-marital
rules; the fraction of countries' ancestors with a nuclear and an extended family structure; the average year the ethnicities were sampled in the \textit{Ethnographic Atlas}. The contemporary controls are: years of civil and interstate conflicts (1816-2007); terrain ruggedness; whether a country was under a communist regime after WWII; the fraction of a country's population
with European descent; oil production per capita; agricultural, manufacturing and services shares of GDP; and the fraction of a country's population who is Catholic, Protestant, other Christian, Muslim, and Hindu. \citet{alesina2013origins} provide the rationale for including each of these controls, and details on how the variables are constructed.

The geo-climatic characteristics included in the IV analysis are: terrain slope, soil depth, average temperature, average precipitation. In the original paper, the geo-climatic characteristics are added linearly, in quadratic forms, and as linear interactions.

\textbf{Details on the DML Analysis.} As in the previous example, the results are obtained with 100 splits and 2-fold cross-fitting. We report median estimates of the coefficients across the splits, and standard errors adjusted for the variability across sample splits using the median method. 
The values of the tuning parameters are the same as in the first example. 

\subsection{The Effect of Skill-Biased Tariff on Growth}
\label{sect:tariff_paper}

\textbf{Details on the Original Analysis.}
For the country-level results, \citet{nunn2010structure} estimate the following regression equation: 

    \begin{equation*}
    \centering
ln\,y_{c1}/y_{c0}= \alpha + \beta_{SB} SB\tau_{c0} + X_{c0} \beta_X + \epsilon_{c},
 \end{equation*}
where $ln\,y_{c1}/y_{c0}$ is the log annual per capita GDP growth in country $c$ between the beginning and the end of the time period considered, $SB\tau_{c0}$ is a measure of initial skill-bias of tariffs, and  $X_{c0}$ represents the controls. 
Three measures of the skill-bias of tariffs are used: the first is the correlation between the industry tariffs and the industry's skill-intensity, while the second and third are based on the difference between the log average tariffs in skill-intensive industries and log average tariffs in unskilled-intensive industries (the two measures differ in the choice of the cut-off value for industry skill-intensity, with the second using a lower cut-off than the third). The controls include: the log of the initial average level of tariffs in the country, three country characteristics measured at the initial period (the log of GDP per capita, the log of human capital, and the log of the ratio of investment-to-GDP), cohort fixed effects (to account for the fact that countries have different initial time periods), region fixed effects (accounting for 10 different regions), and two measures of initial production structure (the log of output in skill-intensive and in unskilled-intensive
industries separately).

Additionally, \citet{nunn2010structure} estimate  the following regression equation, using industry-level data: 
    \begin{equation*}
    \centering
ln\,q_{ic1}/q_{ic0}=\beta_q ln q_{ic0} + \beta_\tau ln \tau_{ic0} + \beta_E ln \bar{\tau}_{c0} + \beta_{SB} SB\tau_{c0} + X_{c0} \beta_X + \alpha_i
+ \epsilon_{ic},
 \end{equation*}
where $ln q_{ic1}/q_{ic0}$ is the average annual log change in
industry output in industry $i$ and country $c$; $ln q_{ic0}$ is the log of industry output in the initial period; $\tau_{ic0}$ is the log initial-period tariff; ln $\bar{\tau}_{c0}$ is the average tariff, $SB\tau_{c0}$ is one of the three measures of skill-bias of tariffs, and $\alpha_i$ are industry fixed effects. The variable $X_{c0}$ indicates the controls which are the same as in the country-level regressions. 

The original results show a strong, positive correlation between skill-biased tariffs and long-term per capita income growth at the country level (Table 4 in \citealp{nunn2010structure}). The correlation is strong also between the skill bias of tariffs and industry output growth, with and without including the initial industry tariff in the regression (Tables 5 and 6 in \citealp{nunn2010structure} respectively). The fact that the size of the coefficient of skill-biased tariffs remains large when adding the variable initial industry tariffs suggests that the mechanism highlighted in the model, i.e. skill-biased tariffs shifting resources towards skill-intensive industries, cannot fully account for the correlation between the treatment variable and long-term growth. \citet{nunn2010structure} further show, with country-level regressions, that the model mechanism can explain up to one quarter of the total correlation between the skill bias of tariffs and long-term growth (Table 7 in the original paper). The paper then investigates other alternative mechanisms that can explain the independent effect of skill-biased tariffs on output growth in Sections V, VI and VII, in the original paper.

\textbf{Details on the DML Analysis.} 
As in the previous examples, the results are obtained with 100 splits and 2-fold cross-fitting. We report median estimates of the coefficients across splits, and standard errors are adjusted for the variability across sample splits using the median method.

The tuning choices are the same as in the previous two examples except for Neural Network in the country-level regressions where the estimates are obtained using 3 neurons and a decay parameter of 0.001.

\subsection{The Effect of Fox News on the Republican Vote Share}
\label{sect:fox_paper}

\textbf{Details on the Original Analysis.}
To produce the main results (see Table IV in the original paper), the authors estimate the following regression:
    \begin{equation*}
    \centering
v_{k,j,2000}^{R,Pres}-v_{k,j,1996}^{R,Pres}= \beta d_{k,2000}^{FOX} +
 \boldsymbol{\Gamma_{2000}}\boldsymbol{X_{k,2000}} + \boldsymbol{\Gamma_{00-90}}\boldsymbol{X_{k,00-90}}+  \boldsymbol{\Gamma_c}\boldsymbol{C_{k,2000}} + \theta_j + \epsilon_{k,j},
\end{equation*}
where $k$ denotes a town in a congressional district $j$. The dependent variable is the change in Republican vote share between the 1996 and the 2000 presidential elections. The treatment variable $d^{FOX}_{k,j,2000}$ is an indicator variable taking the value of 1 for towns where Fox News was available by the year 2000, and 0 otherwise. 
The regression includes demographic controls at the town level: total population, the employment rate, the share of African Americans and of Hispanics, the share of males, the share of the population
with some college education, the share of college graduates, the share of high school graduates, the share of the town that is urban,  the marriage rate, the unemployment rate, and average income. These controls are added both as levels in 2000 ($\boldsymbol{X_{k,2000}}$) and as changes between 1990 and 2000 ($\boldsymbol{X_{k,00-90}}$), and aim at capturing possible confounders that could be correlated with both the availability of Fox News and voting. In addition to the demographic controls, the regression includes a set of cable system features, denoted by $\boldsymbol{C_{k,2000}}$, which are potentially correlated with the treatment variable. These are deciles in the number of channels provided
and in the number of potential subscribers. Finally, fixed effects (congressional district fixed effects or county fixed effects) denoted by $\theta_j$, are added to capture trends in voting that might be common to a geographical area and also correlated with Fox News availability.
In the original analysis, standard errors are clustered at the cable company level. 

The paper also tests whether Fox News increased voter turnout and the Republican vote share in the Senate election. 

The results from the heterogeneity analysis of \citet{dellavigna2007fox} show a negative but insignificant effect for swing district. Additionally, the authors find that the effect of Fox News on the Republican vote share is significantly smaller in towns where the number of cable channels is higher, suggesting a negative impact of higher competition on the effect of Fox News. Moreover, the effect is found to be significantly larger in more urban areas and smaller in more Republican districts. Regarding the latter two findings, the authors point out that in rural areas and in Republican districts the Republican party tends to have a larger vote base to begin with, thus diminishing the share of voters that could potentially be convinced by Fox News.
Out of the four effects, only the differential effect for urban population is significant in both main specifications (county and district fixed effects). The interaction of the treatment variable with the Republican district variable is only significant when including county fixed effects, but not when including district fixed effects, and the opposite is true for the interaction of the treatment with the number of cable channels. The authors also make a note that they find a smaller effect in the South, but this result is not reported in their paper, and we do not focus on it in our analysis.

\textbf{Details on the Analysis with the Causal Random Forest.} 
There are a number of parameters to be set in the causal random forest, algorithm such as the number of trees, the size of the subsample, and the minimum number of control and treatment units in each leaf.  
The \emph{number of trees} is typically chosen as a trade-off between computation times and the test error rate. 
A larger number of trees reduces the Monte Carlo error due to subsampling, which means that the treatment effect predictions will vary less across different forests. A higher number of \emph{minimum treatment and control units} will lead to bigger leaves and a less deep tree. This will predict less heterogeneity. A smaller number will increase the variance as the treatment effect will be estimated with too few observations in a given leaf. Setting a smaller \emph{subsample size} will decrease the dependence across trees, but will increase the variance of each estimate in a tree. The \textit{sizes of the training and estimation samples} are typically fixed to 50$\%$ of the drawn subsample. If there are reasons to allocate more observations to one or the other sample, these proportions can be changed. In the algorithm, there is also a standard parameter for the \textit{number of covariates considered for a split}, before building a tree, within a forest.\footnote{This makes random forests different from bagged trees. In bagged trees the number of predictors considered for a split is equal to the total number of covariates the researcher considers, while in random forests, the number of predictors is strictly less than this total number. The procedure 'decorrelates' the trees (as the trees will be less similar) and the aggregation of predictions across trees will have a lower variance.} This is often set to $\lfloor{K/3}\rfloor$ in the literature, where K is the total number of covariates.

In our analysis, the tuning parameter values are optimised via cross-validation, except the number of trees which is set to 2000. We performed sensitivity analysis with different values for the number of trees (1000 and 5000). The results are available upon request.

In the \textit{cluster-robust} causal forest \citep{athey2019estimating}, when constructing the subsample on which the forest is trained, we do not directly draw observations, but clusters. In addition, in the final step, when constructing the predicted out-of-bag treatment effects, an observation is considered out-of-bag if its cluster was not drawn in the subsample. 

The \textit{variable importance} measure reported in Tables \ref{tab:foxnews_importance_1} and \ref{tab:foxnews_importance_2}
takes into account the proportion of splits over all trees for a particular variable, weighted by depth, and it is useful for describing which covariates influence the most the final estimates when employing the causal forest, as the interpretability of a single tree is lost in this case. 
Recall from the main text that in the causal forest splits are performed if they maximize a criterion function that rewards splits that increase the variance of the treatment effect across leaves, while penalizing splits that increases the variance within a leaf. Hence, higher values for this measure indicate higher importance in terms of heterogeneity of treatment effects.

\subsection{The Effect of Teacher Training on Student Performance}
\label{sect:teacher_paper}

\textbf{Details on the Original Analysis.}
In \citet{loyalka2019does}, the main results are obtained by estimating the following regression equation: 
    \begin{equation*}
    \centering
Y_{i,j}= \alpha_0 + \alpha_1 D_j + X_{ij}\alpha + \tau_k + \epsilon_{i,j},
\end{equation*}
where $Y_{i,j}$ is the outcome, measured at midline or endline, for student \textit{i} in school \textit{j}; $D_j$ is a dummy variable indicating the treatment assignment; the vector $X_{ij}$ includes the control variables, measured at baseline; $\tau_k$ indicates the block fixed effects.\footnote{The schools were randomized within blocks. A block is defined by the year of study the student is enrolled in (i.e. grades 7, 8, or 9), and by the two agencies that implemented the intervention. Hence, the total number of blocks is six.}  The main outcome of interest, student achievement, is measured with a 35-minutes mathematics test at endline. The full set of control variables includes students characteristics (age, gender, parent educational attainment, household wealth), class size, and teacher characteristics (gender, age, experience, education level, rank, a teacher certification dummy, and a dummy indicating whether the teacher majored in math). 

The findings from the heterogeneity analysis suggest that the program has a small positive effect on achievement of students taught by less qualified teachers and a negative effect on students whose teachers are more qualified. In addition, some evidence of heterogeneity is found in terms of whether or not the teachers majored in math, with a negative effect on achievement for those students whose teachers did major in math (this effect is only found when comparing the PD plus follow up with the control group).

\textbf{Details on the Generic ML Analysis.}
In addition to the full set of controls included in the original paper, we add to our analysis the following variables: the baseline values of a number of student-level variables (math self concept, math anxiety, intrinsic motivation for math, instrumental motivation for math, time spent each week studying math), plus a number of variables indicating teachers behaviour in the classroom, evaluated by students at baseline (instructional practices of teacher, teacher care, classroom management of teacher, teacher communication).

The generic ML method takes into account two sources of uncertainty: estimation uncertainty, as the final estimates of interest are obtained conditional on the auxiliary sample, and splitting uncertainty, as the data is randomly split in many auxiliary and main samples. The point estimates are obtained as the median estimates over the different splits of the data. The confidence intervals are constructed by taking the medians of the lower and upper bounds over the random splits. Their nominal level is adjusted to 90\% to account for the splitting uncertainty. In a similar way, the $p$-values are computed based on the median of many random conditional $p$-values, with nominal level adjusted again for splitting uncertainty.

The Best BLP and Best GATES measures are based on maximizing the correlation between the proxy predictor of the conditional average treatment effect, $S(Z)$, and the true conditional treatment effect, $s_{0}(Z)$ (see \citealp{chernozhukov2018generic}).  Table \ref{tab:teacher_comparison} shows that this correlation is the largest for the Neural Network. Therefore, we carry out the HTE analysis using the Neural Network. 

The values of the tuning parameters were optimized via cross-validation for the Elastic Net and Neural Network. For the random forest they are set to default values to save on computation time. For the random forest, the number of trees is set to 2000 and the number of covariates considered for a split is set to $K/3$, which gives a value of 8.

\clearpage
\section{Additional Tables and Figures}

\begin{table}[h!]
\caption{The Effect of Corporate Taxes on Entrepreneurship: Lasso Coefficients of Raw Covariates}
\label{tab:taxes_lasso}
\small
\centering
  \resizebox{\textwidth}{!}{ \begin{tabular}{l c c }
    \toprule
       & (1) & (2)  \tabularnewline

    &  Outcome: & Treatment variable:  \tabularnewline   
   &   Average Entry Rate &First-year Effective Tax Rate  \tabularnewline

     \cmidrule(lr){2-3} 
Log of number of tax payments &	-0,402 &  0,017
 \tabularnewline 
Procedures to start a business & 	-0,006 & 0,001
  \tabularnewline 
Seigniorage 2004 &	-0,003 &  0
 \tabularnewline 
Other taxes &	-0,001 &  0,003
 \tabularnewline 
Rigidity of employment &	0 & 0
  \tabularnewline 
Average inflation (1995-2004) &	0 & 0
  \tabularnewline 
PIT top marginal rate	& 0 &  0,01
 \tabularnewline 
IEF Property Right Index &	0 &  0,001
 \tabularnewline 
VAT and sales tax &	0 &  -0,005
 \tabularnewline 
Tax evasion (GCR) &	0,009 & -0,004
  \tabularnewline 
Log GDP pc 2003 &	0,011 &  -0,02
 \tabularnewline 
EFW Freedom to Trade Internationally Index  &	0,305 &  -0,619
 \tabularnewline 

\bottomrule \tabularnewline
\end{tabular}}
\begin{tablenotes}
\footnotesize
\item \textit{Notes:} The table shows the lasso coefficients of the raw covariates, obtained by estimating the nuisance functions $g(\cdot)$ (column 1) and $m(\cdot)$ (column 2). The lasso coefficients are calculated as the median over splits.
\end{tablenotes}
\end{table}

\begin{sidewaystable}[ph!]
\caption{On the origins of gender roles: Country-level estimates, partially linear model}
\label{tab:ploughcountry}
\small
\centering
  \resizebox{\textwidth}{!}{ \begin{tabular}{l c c c c c c c c}
    \toprule

   & (1) & (2)  & (3) & (4) & (5) & (6)  & (7) & (8)\tabularnewline
     & Lasso & Reg. Tree & Boosting & Forest & Neural Net.  & Ensemble  & Best & OLS \tabularnewline   
  \cmidrule(lr){2-9} 

 \tabularnewline
 & \multicolumn{8}{c}{\textit{Panel A: Female labour force participation 2000}} \tabularnewline \tabularnewline
Traditional plough use & -10.434 & -9.287 & -11.948 & -11.243 & -11.271 & -11.501 & -11.388 & -12.401 \tabularnewline
&(3.195) &(2.817) &(3.129) &(3.181) &(3.227) &(3.182)&(3.18) &(2.964) 
\tabularnewline
   Observations & 165 &165 &165 &165 &165 & 165& 165& 165 \tabularnewline
 \tabularnewline
\hline \tabularnewline
 & \multicolumn{8}{c}{\textit{Panel B: Share of firms with female ownership}} 
 \tabularnewline \tabularnewline
Traditional plough use & -13.168 &-12.316 & -13.114 & -12.769 & -14.556 & -13.287 & -13.54 & -15.241 \tabularnewline
&(3.954) &(3.772) &(3.945) &(4.488) &(3.828) &(4.128)&(4.301) &(4.06) \tabularnewline 
  Observations & 123&123 &123 &123 &123 & 123& 123& 123 \tabularnewline
 \tabularnewline
 \hline \tabularnewline
  & \multicolumn{8}{c}{\textit{Panel C: Share of political positions held by women 2000}} \tabularnewline\tabularnewline
Traditional plough use & -5.196 &-3.617 &-4.29 &-5.41 &-4.48 &-4.817 &-5.029 &-4.821\tabularnewline
  &(1.946) &(1.707) &(1.745) &(1.777) & (1.853)&(1.804) &(1.825)  &(1.782) \tabularnewline
Observations &144 &144 &144 &144 & 144& 144& 144& 144 \tabularnewline \tabularnewline
\hline 
Raw covariates &7 & 7&7 &7 & 7& 7& 7& 7\tabularnewline
\bottomrule \tabularnewline
\end{tabular}}
 \begin{tablenotes}
  \footnotesize
  \item  \textit{Notes:} Analysis of Table 4 (columns 1, 3, 5) of \citet{alesina2013origins} using DML. Column 8 reports the original paper results. Standard errors are reported in parentheses. Standard errors adjusted for variability across splits using the median method are reported for the DML estimates. Robust standard errors are reported in column 8. The number of covariates does not include the treatment variable. 

  \end{tablenotes}
\end{sidewaystable}

\begin{sidewaystable}[ph!]
\caption{The Structure of Tariffs and Long-Term Growth: Industry-level estimates}
\label{tab:tariff2}
\small
\centering
  \resizebox{\textwidth}{!}{ \begin{tabular}{l c c c c c c c c}
    \toprule

   & (1) & (2)  & (3) & (4) & (5) & (6)  & (7) & (8)\tabularnewline
     & Lasso     &  Reg. Tree     &  Boosting     &  Forest     &  Neural Net.     &  Ensemble     &  Best & OLS \tabularnewline   
  \cmidrule(lr){2-9} 

 \tabularnewline
 & \multicolumn{8}{c}{\textit{Panel A: Skill tariff correlation}} \tabularnewline \tabularnewline
Skill tariff correlation& 0.026& 0.019 &0.188 & 0.080& 0.035& 0.153& 0.146 &0.064  \tabularnewline
&(0.053) & (0.072)&(0.103)   &(0.124)& (0.045)& (0.115)& (0.128)&(0.02) 
\tabularnewline

 \tabularnewline
\hline \tabularnewline
 & \multicolumn{8}{c}{\textit{Panel B: Tariff differential (low cut-off)}} 
 \tabularnewline \tabularnewline
Tariff differential (low cut-off) & 0.011 & 0.013 & 0.078 & 0.044& 0.022& 0.058 &0.058 & 0.032 \tabularnewline
 &(0.034)& (0.028)  &(0.042) &(0.071) & (0.022)& (0.072)& (0.078)&(0.01)   \tabularnewline 

\tabularnewline
 \hline \tabularnewline
  & \multicolumn{8}{c}{\textit{Panel C: Tariff differential (high cut-off)}} \tabularnewline\tabularnewline
Tariff differential (high cut-off) & 0.017 & 0.011 &  0.055& 0.050& 0.018& 0.063 &0.058  & 0.040\tabularnewline
&  (0.036)&  (0.031)   & (0.035) & (0.064)& (0.030)&  (0.058)&  (0.065) &(0.009) \tabularnewline
 \tabularnewline 
  \hline
 Observations & 1004 & 1004 & 1004 & 1004 & 1004 &  1004& 1004 &1004 \tabularnewline 
 
Raw covariates &  36 & 36 & 36 & 36 &  36&  36& 36 &36\tabularnewline
\bottomrule \tabularnewline

\end{tabular}}

 \begin{tablenotes}
  \footnotesize
  \item  \textit{Notes:} 
Analysis of Table 5 (columns 1, 2, 4) of \citet{nunn2010structure} using DML. Column (8) reports the original paper estimates. Standard errors are reported in parentheses. Standard errors adjusted for variability across splits using the median method are reported for the DML estimates. Standard errors adjusted for clustering at the country level are reported in column 8. The number of covariates does not include the treatment variable.

  \end{tablenotes}
\end{sidewaystable}

\begin{sidewaystable}[ph!]
\caption{The Structure of Tariffs and Long-Term Growth: Industry-level estimates}
\label{tab:tariff3}
\small
\centering
  \resizebox{\textwidth}{!}{ \begin{tabular}{l c c c c c c c c}
    \toprule

   & (1) & (2)  & (3) & (4) & (5) & (6)  & (7) & (8)\tabularnewline
     & Lasso     &  Reg. Tree     &  Boosting     &  Forest     &  Neural Net.     &  Ensemble     &  Best & OLS \tabularnewline   
  \cmidrule(lr){2-9} 

 \tabularnewline
 & \multicolumn{8}{c}{\textit{Panel A: Skill tariff correlation}} \tabularnewline \tabularnewline
Skill tariff correlation& 0.046& 0.020 &0.164 & 0.086& 0.035& 0.142& 0.103 &0.066 \tabularnewline
&(0.045) & (0.063)&(0.091)   &(0.109)& (0.051)& (0.105)& (0.111)&(0.019) 
\tabularnewline

 \tabularnewline
\hline \tabularnewline
 & \multicolumn{8}{c}{\textit{Panel B: Tariff differential (low cut-off)}} 
 \tabularnewline \tabularnewline
Tariff differential (low cut-off) & 0.026& 0.015 & 0.069 &0.048& 0.022& 0.073 &0.055 & 0.033 \tabularnewline
& (0.024)& (0.033) &(0.037) &(0.061)& (0.026)& (0.059)& (0.059)&(0.01)   \tabularnewline 

\tabularnewline
 \hline \tabularnewline
  & \multicolumn{8}{c}{\textit{Panel C: Tariff differential (high cut-off)}} \tabularnewline\tabularnewline
Tariff differential (high cut-off) &  0.023& 0.013& 0.068& 0.044& 0.019& 0.063 &0.048  & 0.039\tabularnewline
  & (0.021)&  (0.029) &  (0.040)& (0.059) & (0.021)&  (0.051)&  (0.056) &(0.009) \tabularnewline
 \tabularnewline 
  \hline
 Observations & 1004 & 1004 & 1004 & 1004 & 1004 &  1004& 1004 &1004 \tabularnewline 
 
Raw covariates &  37 & 37 & 37 & 37 &  37&  37& 37 &37\tabularnewline
\bottomrule \tabularnewline

\end{tabular}}

 \begin{tablenotes}
  \footnotesize
  \item  \textit{Notes:} 
Analysis of Table 6 (columns 1, 3, 7) of \citet{nunn2010structure} using DML. Column (8) reports the original paper estimates. Standard errors are reported in parentheses. Standard errors adjusted for variability across splits using the median method are reported for the DML estimates. Standard errors adjusted for clustering at the country level are reported in column 8. The number of covariates does not include the treatment variable.
  \end{tablenotes}
\end{sidewaystable}

\begin{table}[t!]
\caption{Fox News - Causal Forest: HTE analysis with district dummies}
\label{tab:foxnews_het2}
\small
\centering
  \resizebox{\textwidth}{!}{\begin{tabular}{l c c c}
    \toprule 
       &  (1) & (2) & (3) \tabularnewline

  & CATE below median & CATE above median  & $p$-value for the difference \tabularnewline

\cmidrule(lr){2-4}

Population, diff. btw. 2000 and 1990 & 0.00413 (0.00242) & 0.00806 (0.00189) & 0.20027 \tabularnewline
\textbf{Share with high school degree, diff. btw. 2000 and 1990} &  \textbf{0.0086 (0.00199)} & \textbf{0.0029 (0.0027)} & \textbf{0.08938} \tabularnewline
Share with some college, diff. btw. 2000 and 1990  & 0.00736 (0.00207) & 0.0039 (0.00227)  & 0.26069 \tabularnewline
Share with college degree,  diff. btw. 2000 and 1990   & 0.00757 (0.00272)& 0.00582 (0.00191) & 0.59872 \tabularnewline
\textbf{Share male,  diff. btw. 2000 and 1990}   & \textbf{0.00949 (0.00222)} & \textbf{0.0035 (0.00231)} &\textbf{0.06126} \tabularnewline
Share African American,  diff. btw. 2000 and 1990   &0.00629 (0.00243) & 0.00666 (0.002)&  0.90674 \tabularnewline
Share Hispanic,  diff. btw. 2000 and 1990   &0.00428 (0.00238) & 0.00737 (0.00208) & 0.32866 \tabularnewline

Unemployment rate, diff. btw. 2000 and 1990  & 0.00366 (0.00238) & 0.00866 (0.00224) & 0.12612 \tabularnewline
Married, diff. btw. 2000 and 1990 & 0.00698 (0.00202) &0.00562 (0.00257) &  0.67592 \tabularnewline
Median income, diff. btw. 2000 and 1990& 0.00628 (0.00224) & 0.00653 (0.0023) & 0.93661 \tabularnewline
Share urban, diff. btw. 2000 and 1990 & 0.00517 (0.00203) & 0.00945 (0.0025)  &  0.18368 \tabularnewline
Population 2000 & 0.00492 (0.00252) & 0.00662 (0.00164) &  0.57185 \tabularnewline

\textbf{Share with some college 2000} & \textbf{0.00328 (0.00204)} & \textbf{0.00964 (0.00249)} &\textbf{0.04809}
\tabularnewline 
Share with college degree 2000  &0.00556 (0.00253) & 0.00679 (0.00185)  & 0.6946  \tabularnewline
Share male 2000 & 0.0055 (0.00194) & 0.00976 (0.00277) &  0.20794 \tabularnewline
Share African American 2000 & 0.0025 (0.00271) & 0.00739 (0.00172) & 0.12759 \tabularnewline
\textbf{Share Hispanic 2000}  & \textbf{0.00136 (0.00225)}  &\textbf{0.00799 (0.00217)}  & \textbf{0.03386} \tabularnewline
Employment rate 2000 & 0.00557 (0.00232) & 0.00771 (0.00215) &  0.50069 \tabularnewline
Unemployment rate 2000 &0.00541 (0.00214)& 0.00741 (0.00235) & 0.52906 \tabularnewline
Share married  2000  &0.00683 (0.00228) & 0.00585 (0.00229) & 0.76121  \tabularnewline
Median income 2000  & 0.00501 (0.00218) & 0.00712 (0.00223) &  0.50006  \tabularnewline
Share urban 2000  & 0.00441 (0.0024) & 0.00673 (0.0019)  & 0.44815  \tabularnewline
No. potential cable subscribers 2000 & 0.00818 (0.00238) & 0.00594 (0.00169) &  0.44436 \tabularnewline
 Decile 1 in no.  potential cable subscribers  & 0.00661 (0.0016) & -0.00787 (0.01626) &  0.37539  \tabularnewline
 Decile 2 in no.  potential cable subscribers  & 0.00664 (0.00165 ) & 0.00084 (0.00861)  & 0.50799 \tabularnewline
 \textbf{Decile 3 in no.  potential cable subscribers}  & \textbf{0.00612 (0.00151)} & \textbf{0.0171 (0.0065)}  & \textbf{0.0999} \tabularnewline
 Decile 4 in no.  potential cable subscribers  & 0.00634 (0.0017) & 0.0077 (0.00393) &  0.75084 \tabularnewline
 Decile 5 in no.  potential cable subscribers  & 0.00667 (0.00174)  &0.00357 (0.00371) &  0.44915 \tabularnewline
 Decile 6 in no.  potential cable subscribers  & 0.00669 (0.00171) & 0.00471 (0.00463) &  0.68762 \tabularnewline
 Decile 7 in no.  potential cable subscribers  & 0.00668 (0.0017) & 0.00531 (0.00492) &  0.79269 \tabularnewline
 \textbf{Decile 8 in no.  potential cable subscribers}  & \textbf{0.00758 (0.00168)} & \textbf{-0.00131 (0.00405)} & \textbf{0.04239} \tabularnewline
 Decile 9 in no.  potential cable subscribers  & 0.0071 (0.00167) & 0.00226 (0.00317) &  0.17685 \tabularnewline
 \textbf{Decile 10 in no.  potential cable subscribers}  & \textbf{0.0045 (0.00207)}  &\textbf{0.01139 (0.00188)} &  \textbf{0.01393} \tabularnewline
 No. cable channels available 2000 & 0.00816 (0.00684) & 0.0065 (0.00148) & 0.812  \tabularnewline
Decile 1 in no.  cable channels available & 0.00645 (0.0017) &  0.00149 (0.02648)  & 0.85167 \tabularnewline
Decile 2 in no.  cable channels available & 0.00655 (0.00153) & 0.01884 (0.0383)  & 0.74845 \tabularnewline
Decile 3 in no.  cable channels available & 0.00657 (0.00161) & 0.00758 (0.01372) &  0.94203 \tabularnewline
Decile 4 in no.  cable channels available& 0.00747 (0.00155) & -0.0101 (0.01332) &  0.18996 \tabularnewline
Decile 5 in no.  cable channels available& 0.00553 (0.00158) & 0.02402 (0.01216) & 0.13149 \tabularnewline
Decile 6 in no.  cable channels available & 0.00569 (0.00164) & 0.01323 (0.00648)  & 0.25923 \tabularnewline
Decile 7 in no.  cable channels available & 0.00585 (0.00192)  &0.00953 (0.00253) & 0.24565 \tabularnewline
Decile 8 in no.  cable channels available & 0.0068 (0.00178) & 0.00355 (0.00398) &  0.45524 \tabularnewline
\textbf{Decile 9 in no.  cable channels available} & \textbf{0.00576 (0.00181)} & \textbf{0.01239 (0.00288)} &  \textbf{0.05169} \tabularnewline

Swing district & 0.00685 (0.00201) & 0.00602 (0.00272)  & 0.80599 \tabularnewline
\textbf{Republican district} & \textbf{0.00693 (0.00187)} & \textbf{0.00084 (0.00264)} &  \textbf{0.06021} \tabularnewline

 \bottomrule 

 \end{tabular}}
        \begin{tablenotes}
  \footnotesize
  \item  \textit{Notes:} The table reports the effect of Fox News on the Republican vote share for towns with values below (column 1) and above (column 2) the median of each variable.  Column 3 presents the $p$-value for the null of no difference between the estimates in columns 1 and 2. Standard errors are reported in parentheses. The estimates are obtained from the causal random forest that includes district dummy variables. As we are not interested in exploring heterogeneity along the congressional districts, the HTE results for district dummy variables are omitted from the table.
\end{tablenotes}
\end{table}

\begin{table}[t!]
\caption{Fox News - Causal Forest: HTE analysis with cluster-robust causal forest}
\label{tab:foxnews_het3}
\small
\centering
  \resizebox{\textwidth}{!}{\begin{tabular}{l c c c}
    \toprule 
       &  (1) & (2) & (3) \tabularnewline

  & CATE below median & CATE above median  & $p$-value for the difference \tabularnewline

\cmidrule(lr){2-4}

Population, diff. btw. 2000 and 1990 & 0.00357 (0.00398)& 0.00829 (0.00299)&  0.34201 \tabularnewline 
Share with high school degree, diff. btw. 2000 and 1990  &0.0088 (0.00311) &0.00225 (0.00323) & 0.14407 \tabularnewline
Share with some college, diff. btw. 2000 and 1990  & 0.00809 (0.00281) &0.00194 (0.00431) & 0.23202 \tabularnewline
Share with college degree, diff. btw. 2000 and 1990   &0.00709 (0.00317) & 0.00604 (0.00329)&  0.8194 \tabularnewline
Share male, diff. btw. 2000 and 1990   & 0.00975 (0.00356) & 0.00308 (0.00268) & 0.13407 \tabularnewline
Share African American, diff. btw. 2000 and 1990   &0.00547 (0.00346) & 0.007 (0.00298) & 0.7364\tabularnewline
Share Hispanic, diff. btw. 2000 and 1990   &0.00369 (0.00383) & 0.00755 (0.00286)& 0.41946 \tabularnewline

Unemployment rate, diff. btw. 2000 and 1990 & 0.00328 (0.00304)& 0.00872 (0.00308)&  0.20834 \tabularnewline
Married, diff. btw. 2000 and 1990 & 0.00622 (0.00327) & 0.00639 (0.00339)&  0.97002 \tabularnewline
Median income, diff. btw. 2000 and 1990&0.0065 (0.00354) & 0.00609 (0.00282)& 0.92735 \tabularnewline
Share urban, diff. btw. 2000 and 1990 & 0.00527 (0.00273) & 0.00881 (0.00372) & 0.44257 \tabularnewline
Population 2000 & 0.00577 (0.00398)  &0.00636 (0.0027) &  0.9022 \tabularnewline

Share with some college 2000 & 0.00532 (0.00321) &0.00785 (0.00376) & 0.60916 \tabularnewline
Share with college degree 2000  &0.00545 (0.00296) & 0.00672 (0.00318) & 0.76975  \tabularnewline
Share male 2000 & 0.00459 (0.00259) & 0.01138 (0.00529) &  0.24942 \tabularnewline
Share African American 2000 & 0.00198 (0.00518) & 0.00731 (0.00265) & 0.35943 \tabularnewline
Share Hispanic 2000  & 0.00071 (0.00378) & 0.00825 (0.00314) & 0.1245 \tabularnewline
Employment rate 2000 &0.0043 (0.00293)  &0.00892 (0.00416) & 0.36452 \tabularnewline
Unemployment rate 2000 & 0.00539 (0.0027) & 0.00728 (0.0035) & 0.66907 \tabularnewline
Share married 2000 & 0.00684 (0.00278) & 0.00561 (0.00355) & 0.78466 \tabularnewline
Median income 2000  & 0.00546 (0.00381)  &0.00648 (0.00272) & 0.82677  \tabularnewline
Share urban 2000  & 0.00534 (0.00404) & 0.00647 (0.00276) & 0.81683 \tabularnewline
No. potential cable subscribers 2000 &0.00744 (0.00616) & 0.00587 (0.00285)&0.81685 \tabularnewline
 Decile 1 in no.  potential cable subscribers  &0.00653 (0.00264) & -0.00486 (0.0162) & 0.48767  \tabularnewline
 Decile 2 in no.  potential cable subscribers  & 0.00655 (0.00268) & 0.00209 (0.0116) & 0.70797\tabularnewline
 Decile 3 in no.  potential cable subscribers  & 0.00594 (0.00258) &0.01893 (0.0111)& 0.25437 \tabularnewline
 Decile 4 in no.  potential cable subscribers  & 0.00628 (0.00256) & 0.00734 (0.00928) & 0.91234 \tabularnewline
 Decile 5 in no.  potential cable subscribers  &0.00677 (0.00273) & 0.00051 (0.00724) & 0.4189 \tabularnewline
 Decile 6 in no.  potential cable subscribers  &0.00691 (0.00278) & 0.00113 (0.00592)& 0.37691 \tabularnewline
 Decile 7 in no.  potential cable subscribers  & 0.00685 (0.00241) &0.00351 (0.01161)& 0.77827\tabularnewline
 Decile 8 in no.  potential cable subscribers  &0.00741 (0.00283) &-0.00051 (0.004) & 0.10608 \tabularnewline
 Decile 9 in no.  potential cable subscribers  & 0.00683 (0.00294)& 0.00274 (0.00409)& 0.41683 \tabularnewline
 Decile 10 in no.  potential cable subscribers  & 0.004 (0.00384) & 0.01147 (0.00369) & 0.16066 \tabularnewline
 No. cable channels available 2000 & 0.00562 (0.00773) & 0.00678 (0.00255)& 0.88643  \tabularnewline
Decile 1 in no.  cable channels available & 0.00646 (0.00274) & -0.00726 (0.03324) & 0.68079 \tabularnewline
Decile 2 in no.  cable channels available & 0.00644 (0.00265) & 0.01768 (0.01293) & 0.39453 \tabularnewline
Decile 3 in no.  cable channels available & 0.00672 (0.00267)  &0.00203 (0.01179) & 0.69811\tabularnewline
Decile 4 in no.  cable channels available& 0.00816 (0.00263) & -0.01645 (0.01506) & 0.10755 \tabularnewline
Decile 5 in no.  cable channels available& 0.00484 (0.00251) & 0.02998 (0.01806) & 0.16774 \tabularnewline
Decile 6 in no.  cable channels available & 0.00537 (0.00289) & 0.0133 (0.0045)& 0.13846 \tabularnewline
Decile 7 in no.  cable channels available & 0.00602 (0.00304) & 0.00867 (0.0042) & 0.60869 \tabularnewline
Decile 8 in no.  cable channels available & 0.00675 (0.00291) & 0.00327 (0.00503) & 0.54915 \tabularnewline
Decile 9 in no.  cable channels available & 0.00543 (0.0027) & 0.0139 (0.00631) & 0.21689 \tabularnewline

Swing district &  0.00634 (0.00308) & 0.00736 (0.00515) & 0.86489\tabularnewline
Republican district &  0.00665 (0.00286) & 0.00079 (0.00604) & 0.38064 \tabularnewline

 \bottomrule 

 \end{tabular}}
        \begin{tablenotes}
  \footnotesize
  \item  \textit{Notes:} The table reports the effect of Fox News on the Republican vote share for towns with values below (column 1) and above (column 2) the median of each variable.  Column 3 presents the $p$-value for the null of no difference between the estimates in columns 1 and 2. Standard errors are reported in parentheses. The estimates are obtained from the cluster-robust causal forest. 
\end{tablenotes}
\end{table}

\begin{table}[h!]
\caption{Fox News - Causal Forest: Variable importance with district dummies}
\label{tab:foxnews_importance_1}
\small
\centering
  \resizebox{.85\textwidth}{!}{\begin{tabular}{l c }
    \toprule 

  Variable & Importance (\%) \tabularnewline
\hline

No. cable channels available 2000 & 6.52 \tabularnewline
No. potential cable subscribers 2000 &  5.23 \tabularnewline

Share employed, diff. btw. 2000 and 1990 &  4.9 \tabularnewline

Share African American 2000 &  4.74 \tabularnewline

Share married  2000  &4.39 \tabularnewline

Unemployment rate, diff. btw. 2000 and 1990   & 4.22\tabularnewline

Decile 10 in no. cable channels 2000 & 4.16 \tabularnewline

Employment rate 2000 &  3.67\tabularnewline

Share with high school degree, diff. btw. 2000 and 1990 &  3.56\tabularnewline

Share with some college 2000 & 3.55\tabularnewline
Population, diff btw. 2000 and 1990 & 3.41 \tabularnewline
Share male, diff btw. 2000 and 1990 &3.38 \tabularnewline
Share Hispanic, diff btw. 2000 and 1990 & 3.27\tabularnewline
Median income, diff btw. 2000 and 1990& 3.25\tabularnewline
Median income 2000 &3.22\tabularnewline
Share Hispanic 2000 &3.21\tabularnewline
Share married, diff btw. 2000 and 1990& 3.07\tabularnewline
Share African American, diff btw. 2000 and 1990& 3.02\tabularnewline
Population 2000 &3.01\tabularnewline
Employment rate 2000& 2.72\tabularnewline
Share with some college, diff btw. 2000 and 1990& 2.67\tabularnewline
Share male 2000 &2.55\tabularnewline
Share with college degree 2000 & 2.49\tabularnewline
Share with college degree, diff btw. 2000 and 1990 & 2.23\tabularnewline
Share with high school 2000 & 2.14\tabularnewline
Decile 10 in no. potential cable subscribers & 2.08\tabularnewline
Share urban population, diff btw. 2000 and 1990& 1.9\tabularnewline
Decile 7 in no. cable channels available& 1.75\tabularnewline
 Decile 9 in no. cable channels available &1.54\tabularnewline
Share of urban population 2000 & 1.4\tabularnewline
Republican district &0.78\tabularnewline
Decile 8 in no. cable channels available & 0.75\tabularnewline
Swing district & 0.74\tabularnewline
Decile 9 in no. potential cable subscribers& 0.36\tabularnewline
Decile 8 in no. potential cable subscribers& 0.07\tabularnewline
Decile 7 in no. potential cable subscribers& 0.02\tabularnewline
 Decile 6 in no. cable channels available  &0.01\tabularnewline
 \bottomrule 
 \end{tabular}}

        \begin{tablenotes}
  \footnotesize
  \item  \textit{Notes:} The table reports the importance of each variable obtained from the causal forest with district dummies. Variables with importance lower than 0.01\% are omitted. 
\end{tablenotes}
\end{table}

\begin{table}[h!]
\caption{Fox News - Causal Forest: Variable importance in cluster-robust causal forest}
\label{tab:foxnews_importance_2}
\small
\centering
  \resizebox{.85\textwidth}{!}{\begin{tabular}{l c }
    \toprule 

  Variable & Importance (\%) \tabularnewline
\hline

No. cable channels available 2000 &  10.4 \tabularnewline

No. potential cable subscribers 2000 &8.22 \tabularnewline

Share with some college 2000 &  4.79 \tabularnewline

Unemployment rate, diff. btw. 2000 and 1990 &  4.35 \tabularnewline

Decile 9 in no. cable channels   & 4.16 \tabularnewline

Decile 10 in no. cable channels   &  4.09\tabularnewline

Employment rate, diff. btw. 2000 and 1990 &  3.86 \tabularnewline

Share African American 2000  & 3.59\tabularnewline

Median income 2000 & 3.39\tabularnewline

Population, diff. btw. 2000 and 1990 & 3.31\tabularnewline

Median income,  diff. btw. 2000 and 1990& 3.1 \tabularnewline
Share married 2000 & 2.9\tabularnewline
Share male,  diff. btw. 2000 and 1990& 2.88\tabularnewline
Decile 7 in no.  cable channels & 2.84\tabularnewline
Unemployment rate 2000  & 2.56\tabularnewline
Share African American, diff. btw. 2000 and 1990  & 2.55\tabularnewline
Share Hispanic, diff. btw. 2000 and 1990  & 2.5\tabularnewline
Share married, diff. btw. 2000 and 1990  & 2.4\tabularnewline
Share Hispanic 2000  & 2.38\tabularnewline
Share with high school degree, diff. btw. 2000 and 1990  & 2.28\tabularnewline
Share urban, diff. btw. 2000 and 1990  & 2.23\tabularnewline
Share male 2000  & 2.2\tabularnewline
Decile 10 in no. potential cable subscribers & 2.19\tabularnewline
Population 2000  & 2.14\tabularnewline
Share with some college, diff. btw. 2000 and 1990  & 2.08\tabularnewline
Share with college degree 2000 & 1.98\tabularnewline
Employment rate 2000  & 1.79\tabularnewline
Share with college degree, diff. btw. 2000 and 1990  & 1.58\tabularnewline
Share with high school degree 2000  & 1.57\tabularnewline
Share urban 2000  & 1.55\tabularnewline
Republican district  & 1.37\tabularnewline
Swing district  & 0.98\tabularnewline
Decile 8 in no. cable channels  & 0.97\tabularnewline
Decile 9 in no. potential cable subscribers & 0.51\tabularnewline
Decile 8 in no. potential cable subscribers & 0.21\tabularnewline
Decile 7 in no. potential cable subscribers & 0.06\tabularnewline
Decile 6 in no. cable channels  & 0.03\tabularnewline
Decile 6 in no. potential cable subscribers& 0.02\tabularnewline
 \bottomrule 
 \end{tabular}}

        \begin{tablenotes}
  \footnotesize
  \item  \textit{Notes:} The table reports the importance of each variable obtained from the cluster-robust causal forest. Variables with importance lower than 0.01\% are omitted. 
\end{tablenotes}
\end{table}

\begin{table}[t!]
\caption{Teacher Training - Generic Method: Comparison of ML methods}
\label{tab:teacher_comparison}
\small
\centering
  \resizebox{.7\textwidth}{!}{\begin{tabular}{l c c c}
    \toprule 
   & (1) & (2)  & (3) \tabularnewline
  & Elastic net & Neural network & Random forest   \tabularnewline
\cmidrule(lr){2-4}

 \vspace{2mm}
Best BLP & 0.012& 0.014& 0.011 \tabularnewline  \vspace{3mm}
Best GATES & 0.115& 0.121& 0.099 \tabularnewline  
 \bottomrule 
 \end{tabular}}
 \vspace{2mm}
        \begin{tablenotes}
  \footnotesize
  \item  \textit{Notes:} The table compares the performance of the three ML methods used to produce the proxy predictors. The performance measures Best BLP and Best GATES are computed as medians over 100 splits.
\end{tablenotes}
\end{table}

\begin{table}[h!]
\caption{Teacher Training - Generic Method: GATES of most and least affected groups}
\label{tab:teacher_gates}
\small
\centering
  \resizebox{\textwidth}{!}{\begin{tabular}{l c c c}
    \toprule 
 &  (1) & (2) & (3) \tabularnewline
& 20\% most affected &  20\% least affected & Difference \tabularnewline
\cmidrule(lr){2-4}

 \vspace{2mm}
Effect of teacher training on student achievement & 0.164 & -0.179 & 0.365
\tabularnewline

 \vspace{2mm}
90\% Confidence Interval &(0.048,0.279) &(-0.301,-0.058) &(0.198,0.533) \tabularnewline  
 \vspace{2mm}
$p$-value & 0.011 & 0.092 & 0.001 \tabularnewline
 \bottomrule 
 \end{tabular}}
        \begin{tablenotes}
  \footnotesize
  \item  \textit{Notes:}  The estimates are obtained using neural network to produce the proxy predictor \textit{S(Z)}. The values reported correspond to the medians over 100 splits.
\end{tablenotes}
\end{table}

\begin{table}[h!]
\caption{Teacher Training - Generic Method: Classification Analysis}
\label{tab:teacher_clan2}
\small
\centering
  \resizebox{\textwidth}{!}{\begin{tabular}{l c c c}
    \toprule 
   & (1) & (2) & (3) \tabularnewline
  & 20\% most affected & 20\% least affected & \textit{p}-value for the difference \tabularnewline
\cmidrule(lr){2-4}

Baseline instructional practices of teacher & 0.211 &0.053&  0.000\tabularnewline
& (0.149,0.272) & (-0.009,0.114)\tabularnewline
Teacher's baseline classroom management &0.065 &0.074 & 0.000\tabularnewline
& (0.003,0.127) & (0.012,0.136)\tabularnewline
Teacher gender &0.529& 0.536 & 1.000\tabularnewline
& (0.497,0.562) & (0.504,0.568)\tabularnewline
Teacher certification dummy & 0.970 &1.000  & 0.000 \tabularnewline

&(0.962,0.977)& (0.992,1.008)\tabularnewline
Student's baseline instrumental motivation for math & -0.111& 0.224&  0.000 \tabularnewline
& (-0.173,-0.050) & (0.162,0.285)\tabularnewline
Student's baseline time spent each week studying math &-0.073& 0.204  & 0.000\tabularnewline
& (-0.142,-0.004) & (0.135,0.273)\tabularnewline
Student's baseline math self-concept & -0.380& 0.317 & 0.000\tabularnewline
& (-0.441,-0.319)& (0.256,0.378)\tabularnewline
Teacher majored in math & 0.309& 0.507  & 0.000\tabularnewline
& (0.278,0.340) &(0.475,0.538)\tabularnewline
Mother education level & 0.570& 0.425  &0.000 \tabularnewline
& (0.538,0.602)& (0.393,0.457)\tabularnewline
Teacher's baseline communication & -0.088 &0.251  &0.000 \tabularnewline
& (-0.152,-0.025)& (0.187,0.314)\tabularnewline
Student's baseline intrinsic motivation for math
 &-0.294& 0.299  &0.000 \tabularnewline
& (-0.356,-0.232) &(0.237,0.362)\tabularnewline
Baseline teacher care & -0.165 &0.311 &0.000 \tabularnewline
& (-0.228,-0.103)& (0.249,0.374)\tabularnewline
Household asset index & -0.421& 0.240  &0.000 \tabularnewline
& (-0.491,-0.351) & (0.170,0.310)\tabularnewline
Father education level & 0.583& 0.589&1.000 \tabularnewline
& (0.551,0.614)& (0.557,0.621)\tabularnewline

\bottomrule 

 \end{tabular}}
        \begin{tablenotes}
  \footnotesize
  \item  \textit{Notes:} The table shows the average value of the teacher and student characteristics for the most and least affected groups.  The estimates are obtained using neural network to produce the proxy predictor \textit{S(Z)}. Confidence intervals with 90\% nominal level are reported in parenthesis. All variables, except \textit{Teacher gender, Teacher certification dummy, Teacher majored in math, Mother education level} and \textit{Father education level} are normalized. The values reported correspond to the medians over 100 splits.
\end{tablenotes}
\end{table}

\begin{table}[h!]
\caption{Teacher Training - Generic Method: Correlation of the covariates with $S(Z)$}
\label{tab:teacher_corr}
\small
\centering
  \resizebox{.85\textwidth}{!}{\begin{tabular}{l c }
    \toprule 

  Variable & Correlation \tabularnewline
\hline

Teacher college degree & -0.237\tabularnewline

Teacher training hours & 0.125\tabularnewline
Teacher ranking & 0.116 \tabularnewline
Student age & 0.111\tabularnewline
Teacher experience (years) & 0.101\tabularnewline
Student female &-0.094\tabularnewline
Teacher age & 0.089\tabularnewline
Math score at baseline (normalized) & 0.075\tabularnewline
Student baseline math anxiety & 0.063\tabularnewline
Class size & -0.060\tabularnewline
Baseline instructional practices of teacher & 0.053\tabularnewline
Teacher’s baseline classroom management &0.051\tabularnewline
Teacher gender & -0.045\tabularnewline
Teacher certification dummy & 0.036\tabularnewline
Student's baseline instrumental motivation for math  & 0.025\tabularnewline
Student's baseline time spent each week studying math & 0.022\tabularnewline
Student's baseline math self-concept & -0.021\tabularnewline
Teacher majored in math &-0.016\tabularnewline
Mother education level & 0.009\tabularnewline
Teacher's baseline communication & -0.008\tabularnewline
Student's baseline intrinsic motivation for math & 0.008\tabularnewline
Baseline teacher care & -0.006\tabularnewline
Household asset index & -0.005\tabularnewline
Father education level & -0.003\tabularnewline
 \bottomrule 
 \end{tabular}}

        \begin{tablenotes}
  \footnotesize
  \item  \textit{Notes:} The table reports the correlation of each covariate with the proxy predictor $S(Z)$.
\end{tablenotes}
\end{table}

\clearpage 
\setcounter{figure}{0}
\renewcommand\thefigure{\thesection.\arabic{figure}}

\clearpage

 \begin{figure}[!h]
  \caption{Corporate Taxes on Entrepreneurship: Lasso Coefficients of Interaction Terms}
\label{fig:taxlasso}
\centering
\begin{minipage}[c]{.8\textwidth} 
\centering

  \includegraphics[width=\columnwidth]{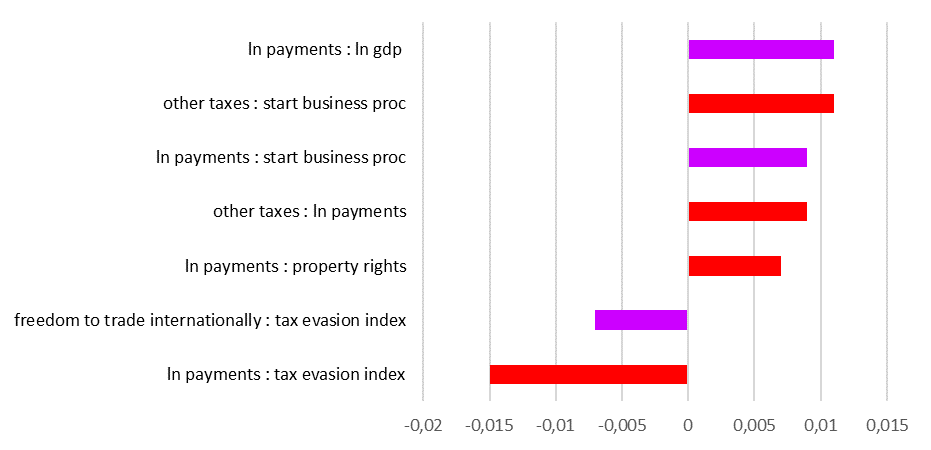} 
  \caption*{\small{Figure 2.1: $\hat{m}(\cdot)$}}

  \end{minipage}\hfill{}
  \begin{minipage}[c]{.8\textwidth} 
  \centering

   \includegraphics[width=\columnwidth]{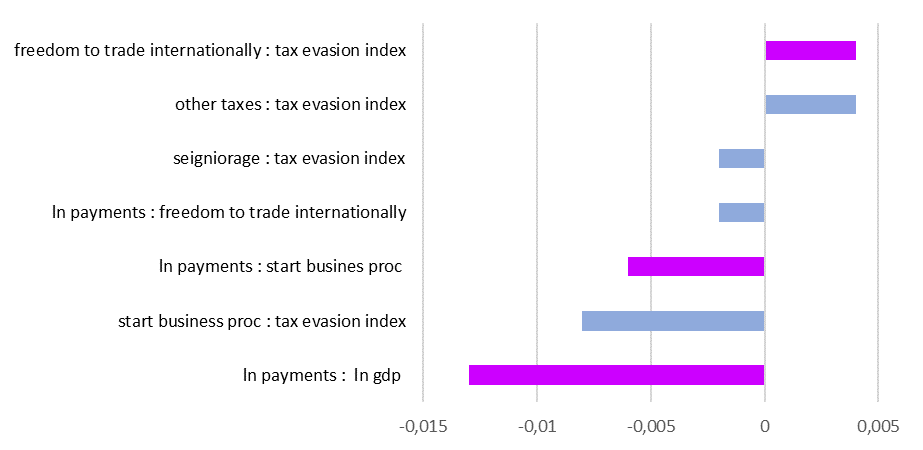} 
   \caption*{\small{Figure 2.2: $\hat{g}(\cdot)$}}
   \end{minipage}
      \begin{tablenotes}
  \footnotesize
  \item  \textit{Notes:} The figure plots the seven largest lasso coefficients of the interaction terms, obtained estimating the nuisance functions $m(\cdot)$ and  $g(\cdot)$. Colons indicate interactions of variables. The treatment variable D, is the first year effective corporate tax rate. The dependent variable Y is the average entry rate.
  The lasso coefficients are calculated as the median over splits. 
\end{tablenotes}
\end{figure}

\end{document}